\documentclass[sn-mathphys]{sn-jnl}

\jyear{2021}%

\theoremstyle{thmstyleone}%
\theoremstyle{thmstyletwo}%

\theoremstyle{thmstylethree}%

\usepackage{lineno}

\begin{document}

\title[Neural Downscaling]{Neural Downscaling for Complex Systems: from Large-scale to Small-scale by Neural Operator}


\author[1]{\fnm{Pengyu} \sur{Lai}}\email{laipengyu@sjtu.edu.cn}

\author[1]{\fnm{Jing} \sur{Wang}}

\author[1]{\fnm{Rui} \sur{Wang}}

\author[1]{\fnm{Dewu} \sur{Yang}}

\author[1]{\fnm{Haoqi} \sur{Fei}}

\author*[1]{\fnm{Hui} \sur{Xu}}\email{dr.hxu@sjtu.edu.cn}

\affil*[1]{\orgdiv{School of Aeronautics and Astronautics}, \orgname{Shanghai Jiao Tong University}, \orgaddress{\city{Shanghai}, \postcode{200240}, \country{China}}}


\abstract{
Predicting and understanding the chaotic dynamics in complex systems is essential in various applications. However, conventional approaches, whether full-scale simulations or small-scale omissions, fail to offer a comprehensive solution. This instigates exploration into whether modeling or omitting small-scale dynamics could benefit from the well-captured large-scale dynamics.
In this paper, we introduce a novel methodology called Neural Downscaling (ND), which integrates neural operator techniques with the principles of inertial manifold and nonlinear Galerkin theory. ND effectively infers small-scale dynamics within a complementary subspace from corresponding large-scale dynamics well-represented in a low-dimensional space. The effectiveness and generalization of the method are demonstrated on the complex systems governed by the Kuramoto-Sivashinsky and Navier-Stokes equations. As the first comprehensive deterministic model targeting small-scale dynamics, ND sheds light on the intricate spatiotemporal nonlinear dynamics of complex systems, revealing how small-scale dynamics are intricately linked with and influenced by large-scale dynamics. 
}



\maketitle

\section{Main}\label{sec1}
In the realm of science and engineering, 
the studies of complex systems that exhibit dynamics spanning multiple spatiotemporal scales
dominates the most important advances and designs~\cite{temam2012infinite,strogatz2000}.
Chaotic behaviors are commonly observed in spatiotemporal systems, characterized by extreme sensitivity to initial conditions and the absence of long-term predictability.
Predicting and understanding the chaotic dynamics in complex systems is essential for various applications, including weather modelling and forecasting\cite{bi2023accurate,lam2023learning}, ecological modelling\cite{blasius1999complex}, and understanding the behaviour of turbulent flows\cite{mukherjee2023intermittency}. 
However, these systems often exhibit intricate patterns and interactions across multiple spatiotemporal scales, making their prediction and analysis challenging\cite{holmes2012turbulence}.


To tackle the complexity of these systems, \textbf{full-scale simulation} have been employed through directly solving the original dynamical equations that govern the corresponding complex system without additional assumptions. 
In the context of fluid mechanics known for turbulence's extreme complexity\cite{rogallo1984numerical}, continuum fluid dynamics usually can be described by Navier-Stokes equations that are numerically solved by various numerical schemes such as finite element method, finite volume method\cite{moin1998direct}, and rarefied fluid dynamics by Boltzmann equation numerically solved by lattice Boltzmann method\cite{succi2001lattice}. Although these methods can obtain a high-fidelity system description, their computational cost and time are usually not affordable for scenarios representing both natural and engineering fluid flows. 
Apart from these traditional methods, machine learning (ML) emerges as a novel and powerful tool for enhancing the modelling capability\cite{kutz2017deep}. However, the ML-assisted general full-scale simulations are still unrealistic due to the expensive generation of training data, poor generalization on limited data and difficulty in learning multiple scales.

To enhance the small-scale simulations, \textbf{small-scale (unresolved) scale modelling simulation} has also been developed, resulting in a more computationally feasible approach. This method tackles surrogate dynamic equations, retaining large-scale dynamics (the dominant patterns and structures within the system) while modelling the small-scale dynamics (the intricate particulars and variability) through an efficient method, often relying on empiricism or assumptions.  
For instance, the Mori-Zwanzig formalism\cite{sture1974strategies} and the extended evolutionary renormalization group (RG) method\cite{schmuch2013} treat small-scale dynamics as stochastic noise. In the turbulence community, classic examples include the sub-grid scale model for Large-Eddy Simulation\cite{smagorinsky1963general} and one-equation turbulence model for Reynolds-Averaged Navier Stokes equations\cite{spalart1992one}, which represent microscopic dynamics as Reynolds stresses. Extensive research has also been dedicated to improving these surrogate models, aiming for more sophisticated and accurate models for small-scale scale dynamics using machine learning techniques and adequate data\cite{kutz2017deep}. These methods can be categorized into explicit methods, where ML-assisted closure models directly substitute surrogate models\cite{duraisamy2019turbulence}, and implicit methods, where large-scale dynamics are mapped to full-scale dynamics by ML models such as super-resolution techniques\cite{fukami2019super}. These approaches achieve computational efficiency by treating small-scale scale dynamics as a simplified model and taking a proper trade-off between precision and cost. 
However, these model construction methodologies inherently limit their ability to offer detailed and fast representation of small-scale dynamics.
Generic approaches often depend on stochastic or implicit methods, while deterministic and explicit models are typically applicable only to specific systems (e.g., turbulence models cannot be universally generalized).

To further enhance the efficiency, reduced-order models have been employed to conduct the \textbf{small-scale omission simulation}, which simulates dynamics at the small scale by appropriately omitting finer details (such as truncating high-order terms). Common techniques include adiabatic elimination\cite{van1985elimination}, classical center-manifold theory\cite{mercer1990centre}, and inertial manifold theory\cite{foias1988inertial}. 
Advanced data-driven approaches facilitate the development of reduced-order models capable of accurately capturing essential dynamics, interactions, and patterns within dynamic data\cite{schmuch2013,lusch2018,cenedese2022}. 
However, the accuracy of long-term dynamics predicted by these models heavily relies on their methodology of complexity reduction. Excessive complexity reduction neglects
essential information contained in the unresolved degrees and degenerates the accuracy, while deficient complexity reduction intensifies the computational cost. An appropriate scale omission is the central question for reduced-order models.  
Moreover, \textbf{qualitative simulation} has been applied to solve the simple dynamical models that are not derived quantitatively from original dynamic equations but are developed to capture the key features in a complex system by mimicking the key physical processes\cite{majda2014conceptual, farazmand2017variational}. 
These methods prioritize capturing significant statistical features of vastly more complex systems qualitatively, thus offering substantial computational savings. However, they inherently do not reproduce detailed dynamics within spatial and temporal regions, limiting their ability to explore pattern and scale interactions within complex systems.

Full-scale simulation poses significant challenges with current technology, prompting researchers in dynamic systems to focus on small-scale modeling and small-scale omission simulation due to the industry's need for rapid simulation and its relative simplicity. 
Considering the wide accessibility and accurate representation of large-scale dynamics, this leads to a crucial and open issue: whether the modeling or omission of small-scale dynamics could benefit from the well-represented large-scale dynamics. This mirrors challenges in addressing the causal emergence problem\cite{barnett2023dynamical, hoel2013quantifying} and self-organized criticality problem\cite{bak1988self}. From the viewpoint of the inertial manifold, one can treat small-scale dynamics as a slaved system that responds to the slow evolution of their corresponding large-scale dynamics, providing a way to build the theoretical guarantee of small-scale dynamics modelling exclusively based on large-scale dynamics.
Computationally, the approximation of inertial manifold involves the nonlinear Galerkin method, where modal expansion coefficients are distributed randomly in mode order and involve a wide range of spatial scales and it is difficult to find a single time scale for the evolution of each mode, especially for the higher mode numbers. Therefore, neural network is considered a powerful tool for determining the modal expansion coefficients to establish a high-fidelity mapping.
Consequently, this fusion of inertial manifold theory and neural networks introduces a novel methodology termed Neural Downscaling (ND), which tackles the critical challenge in an explicit and deterministic manner, capable of generalizing to arbitrary complex systems, while offering a fresh perspective on the underlying mechanics of complex systems. 
Moreover, ND holds promise for resolving issues related to low-fidelity dynamics
and advancing our understanding of complex systems. Its potential applications extend to diverse fields such as emergent abilities in large language models\cite{wei2022emergent}, consciousness generation in brain\cite{tononi2016integrated} and the emergence in causality\cite{hoel2013quantifying}.

\section{Results}\label{sec2}
We consider the following classical form of nonlinear dynamic system:
\begin{equation}
 \frac{d u(t)}{d t}= \mathscr{A}u(t)+f,
\end{equation}
where $\mathscr{A}=\mathscr{L}+\mathscr{N}$ is a differential operator which consists of a linear part $\mathscr{L}$ and a nonlinear part $\mathscr{N}$ and $f$ is a source term.    
To find the solution $u(t)$ in the functional space $\mathscr{X}$, consider a decomposition $\mathscr{X}=\mathscr{X}^M \oplus \mathscr{X}^{M^{\perp}}$, where $\mathscr{X}^M$ is defined in the $M \in \mathbb{N^+}$ dimension functional space and $\mathscr{X}^{M^{\perp}}$ is defined in the corresponding complementary functional space. 
In view of the nonlinear Galerkin method\cite{foias1988inertial,marion1989nonlinear}, the nonlinear dynamic system can be decomposed into:
\begin{subequations}\label{eq:inertial}
\begin{equation}\label{eq:inertial2a}
 \frac{d p_m(t)}{d t}=\mathscr{L}p_m(t) +\mathscr{P}_M\mathscr{N}\left(p_m(t)+q_m(t)\right)+\mathscr{P}_Mf,
 \end{equation}
 \begin{equation}\label{eq:inertial2b}
 \frac{d q_m(t)}{d t}=\mathscr{L}q_m(t) +\mathscr{Q}_M\mathscr{N}\left(p_m(t)+q_m(t)\right)+\mathscr{Q}_Mf ,
 \end{equation}
\end{subequations}
where $\mathscr{P}_M$ and $\mathscr{Q}_M$ are the projection operators defined on $ \mathscr{X}^M$ and $\mathscr{X}^{M^{\perp}}$, respectively.
$p_m(t) = \mathscr{P}_M u(t) \in \mathscr{X}^M$ and $q_m(t) = \mathscr{Q}_M u(t) \in\mathscr{X}^{M^{\perp}}$ are the projected solutions. Typically, $p_m$ is of comparable magnitude to $u$, while $q_m$ is significantly smaller in comparison to $p$ and $u$. Accordingly, $p_m(t)$ denotes the large-scale decomposition dominating the evolution of dynamic system, while $q_m(t)$ denotes the small-scale one slaved by $p_m(t)$\cite{igo2020}. From the perspective of the RG methods\cite{chen1994},  the decomposition in Eq.\ref{eq:inertial} into $p_m(t)$ and  $q_m(t)$ can be regarded as slow and fast modes, respectively.  

Since $q_m$ is small, the nonlinear parts involved $q_m$, i.e. $\mathscr{N}(q_m, q_m)$, $\mathscr{N}(q_m, p_m)$ and $\mathscr{N}(p_m, q_m)$, are also small in comparison with $\mathscr{N}(p_m, p_m)=\mathscr{N}p_m$, then they can be neglected reasonably. If the relaxation time of the linear part of Eq. \ref{eq:inertial2b} is much smaller than that of Eq.\ref{eq:inertial2a}, an acceptable approximation can be obtained by Eq. \ref{eq:inertial2b}:
\begin{equation}\label{eq3}
\mathscr{L}q_m(t)+\mathscr{Q}_M\mathscr{N}p_m(t)+\mathscr{Q}_Mf=0.
\end{equation}
This finally leads us to establish a mapping from large-scale dynamic $p_m$ to small-scale dynamic $q_m$ in a manifold $\mathscr{X}$:
\begin{equation}\label{eq4}
\begin{aligned}
    q_m(t)=&\mathscr{L}^{-1}(-\mathscr{Q}_M\mathscr{N}p_m(t)-\mathscr{Q}_Mf)\\
&p_m \in \mathscr{X}^{M}, \quad q_m \in \mathscr{X}^{M^{\perp}}
\end{aligned}
\end{equation}
Mathematically, our objective is to construct a mapping $\Phi: \mathscr{X}^M\rightarrow \mathscr{X}^{M^{\perp}}$ such that $q_m(t)$ is determined by $\Phi(p_m(t))$\cite{foias1988inertial}. 
In this paper, we consider the following abstract description of neural network (neural operator in the functional sense):
\begin{equation}
\mathcal{G}_{\varrho}(\omega ; \theta): \mathscr{X}^M \times \mathbb{R}^{P(K)} \rightarrow \mathscr{X}^{M^{\perp}},
\end{equation}
where $\varrho$ denotes the activation function, $\omega$ represents the input of neural operator, $\theta$ indicates the total parameter set of weights (with the omission of biases for brevity). $K$ is the layer-level weights number tensor excluding input layer. $P(K)$ is the total parameter number of weights and equal to $\vert \theta \vert$ (cardinality). In the above definition, the layer number is not given explicitly, which depends on the architecture $(K, \varrho)$ of neural operator. 

\begin{figure}[hbtp]
  \centering
  \includegraphics[width=0.95\textwidth]{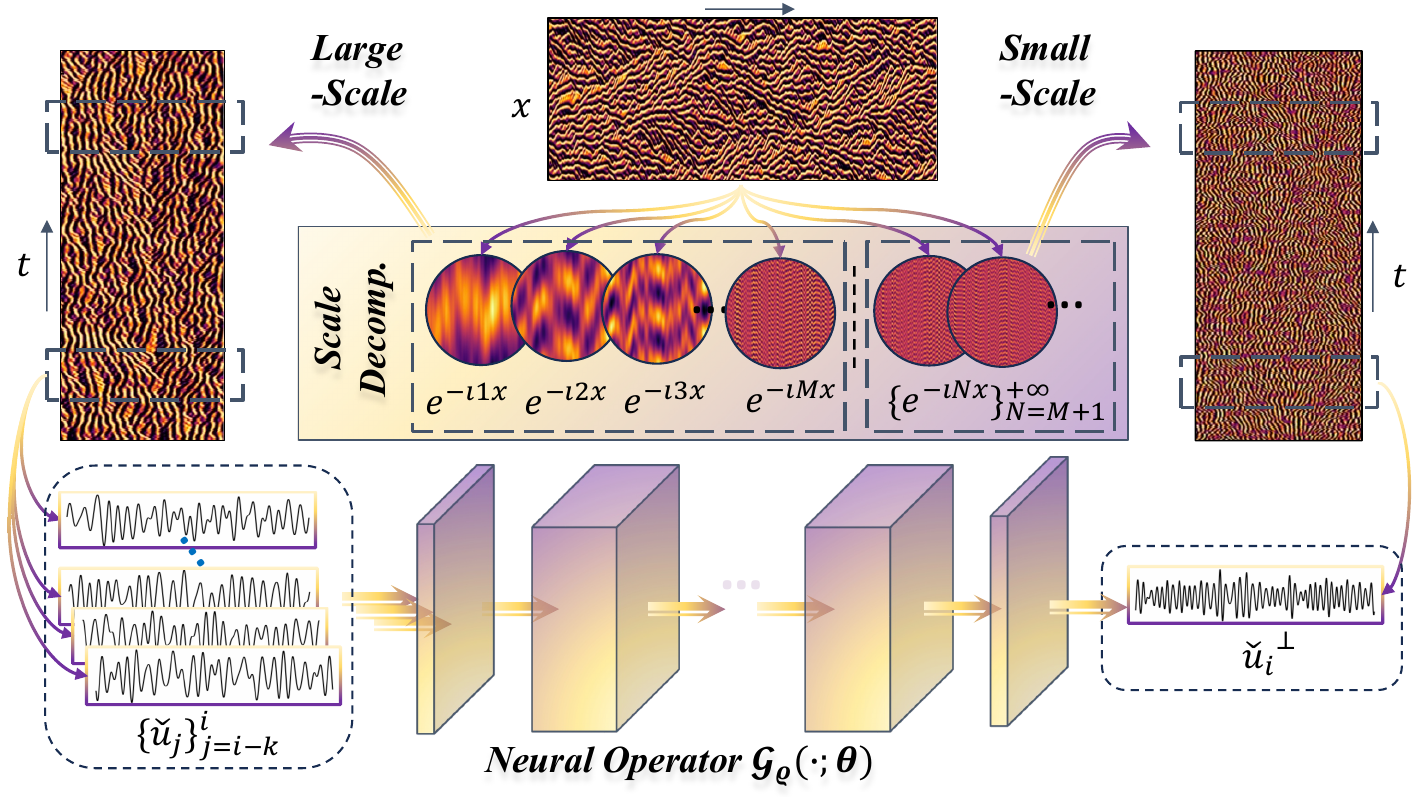}
  \caption{The illustrative sketch of the graph mapping from large-scale dynamics $\breve{v}_i$ on $\mathscr{X}^M$ to slaved small-scale dynamics $\breve{v}^{\perp}_j$ on $\mathscr{X}^{M^{\perp}}$. The graph is approximated by the convolutional neural network, where temporal-dependent input $\{\breve{v}_j\}^i_{j=i-k}$ is considered. The original dataset is formed by a temporal sequence sampled from a dynamics system, and then scale decomposition on the dataset is achieved by the Fourier decomposition method, resulting in a temporal dataset of large-scale dynamics and  a temporal dataset of slaved small-scale dynamics.}
  \label{Network structure} 
\end{figure}

Currently,  we are considering approximating $\Phi$ by $\mathcal{G}_{\varrho}(\omega ; \theta)$ as shown in Fig. \ref{Network structure}.  For convenience,  let $\breve{u}_i$ indicate the $u_i$ projected onto $\mathscr{X}^{M}$ and the corresponding $\breve{u}_i^{\perp}$ onto $\mathscr{X}^{M^{\perp}}$. To represent the slaved small-scale physics by the large-scale physics, the relative loss for the neural operator is defined as:
\begin{equation}\label{eq:loss}
\mathscr{J}_k(\theta)=\frac{\underset{\breve{ u}_i\oplus\breve{ u}_{i}^{\perp}\in\mathscr{U}}{\mathbb{E}}\left\|\mathcal{G}_{\varrho}\left(\{\breve{ u}_j\}_{j=i-t}^{i}; \theta\right)-\breve{ u}_{i}^{\perp}\right\|^2_p}{\underset{\breve{ u}_i\oplus\breve{ u}_{i}^{\perp}\in\mathscr{U}}{\mathbb{E}}\left\|\breve{ u}_{i}^{\perp}\right\|^2_p},
\end{equation}
where the set $\mathscr{U}$  denotes the training subset in $\mathscr{X}$, $t$ indicates continuous previous steps and $p \in \mathbb{R^+}$ is the order of $l^p$-norm. In this study, the classical convolutional neural operator is employed to establish the approximate mapping $\tilde{\Phi}=\mathcal{G}_{\varrho}(\cdot ; \theta)$, which is referred to as ND. Temporal-dependent input is based on the idea that a precise representation of the small-scale dynamics necessitates incorporating historical large-scale data. This concept draws inspiration from the integro-differential operator of the small-scale dynamics\cite{foias1988inertial}. Detailed definition of neural operator refers to \nameref{Neural operator} in \nameref{Methods}.  In this paper, Euclidean norm $p=2$ is utilized for training ND.

\subsection{Kuramoto-Sivashinsky equation}
The Kuramoto-Sivashinsky (KS) equation is a nonlinear partial differential equation that arises in various physical contexts, such as flame propagation, reaction-diffusion systems, fluid films, and plasma dynamics. It is known for its rich and complex spatiotemporal behaviour, exhibiting patterns such as cellular structures, travelling waves, and chaos. As an illustrative instance of a spatiotemporal chaotic system, the KS equation is first considered:
\begin{equation}\label{eq1}
    \begin{aligned}
v_t+v_{x x}+v_{x x x x}+ vv_x=0,& \quad  \text{in} \  \mathcal{P} \times [0,T] \ ,\\
v(x, 0)=v_0(x),& \quad  \text{in} \  \mathcal{P} \ ,
\end{aligned}
\end{equation}
where $\mathcal{P}:=]0,L[$ indicates the periodic domain, $v_0(x)$ denotes the initial condition. 
It has been demonstrated that solutions of the KS equation are exponentially attracted toward a finite-dimensional inertial manifold in phase space\cite{temam2012infinite}. 
These manifolds have emerged as a rich source of nonlinear dynamics, sharing dynamic similarities with classical spatiotemporal systems governed by the Navier–Stokes (NS) equations.

To generate the numerical solution as the dataset, the KS equation is solved by the pseudospectral method combined with the exponential time-differencing fourth-order Runge–Kutta formula, generating a $N$-step temporal sequence $\{v_i\}_{i=1}^{N}$. The initial condition is set to $v_0=cos(16 \cdot 2\pi x/L)$. With $L = 64 \pi$, the time integration is implemented until the final time $T= 3000$ where $1024$ Fourier modes are employed to discretize the spatial domain. 
A total of $N=11000$ continuous snapshots are generated, starting from $250s$ when chaos is fully developed. In order to obtain the approximate mapping $\mathcal{G}_{\varrho}(\cdot ; \theta)$ between the large-scale dynamic in $\mathscr{X}^{M}$ and the slaved small-scale dynamic in $\mathscr{X}^{M^{\perp}}$,  the first $4000$ continuous snapshots are adopted to form the training dataset and others to form the test dataset. The scale decomposition parameter $M = 33$ is determined where about $80\%$ of the energy lies in the space $\mathscr{X}^M$. Through scale decomposition, projected temporal sequences $\{\breve{v}_i\}_{i=1}^N$ on $\mathscr{X}^{M}$ and $\{\breve{v}_i^{\perp}\}_{i=1}^N$ on $\mathscr{X}^{M^{\perp}}$ are obtained. To perform the training of ND, the preceding 4 continuous snapshots, denoted as $\{\breve{v}_j\}^{i-1}_{j=i-4}$, along with the current snapshot $\breve{v}_i$ collectively constitute a temporal sequence $\{\breve{v}_j\}^i_{j=i-4}$ as the input, and $\breve{v}_i^{\perp}$ as the target. 

\begin{figure}[htbp]
  \centering
  \includegraphics[width=0.8\textwidth]{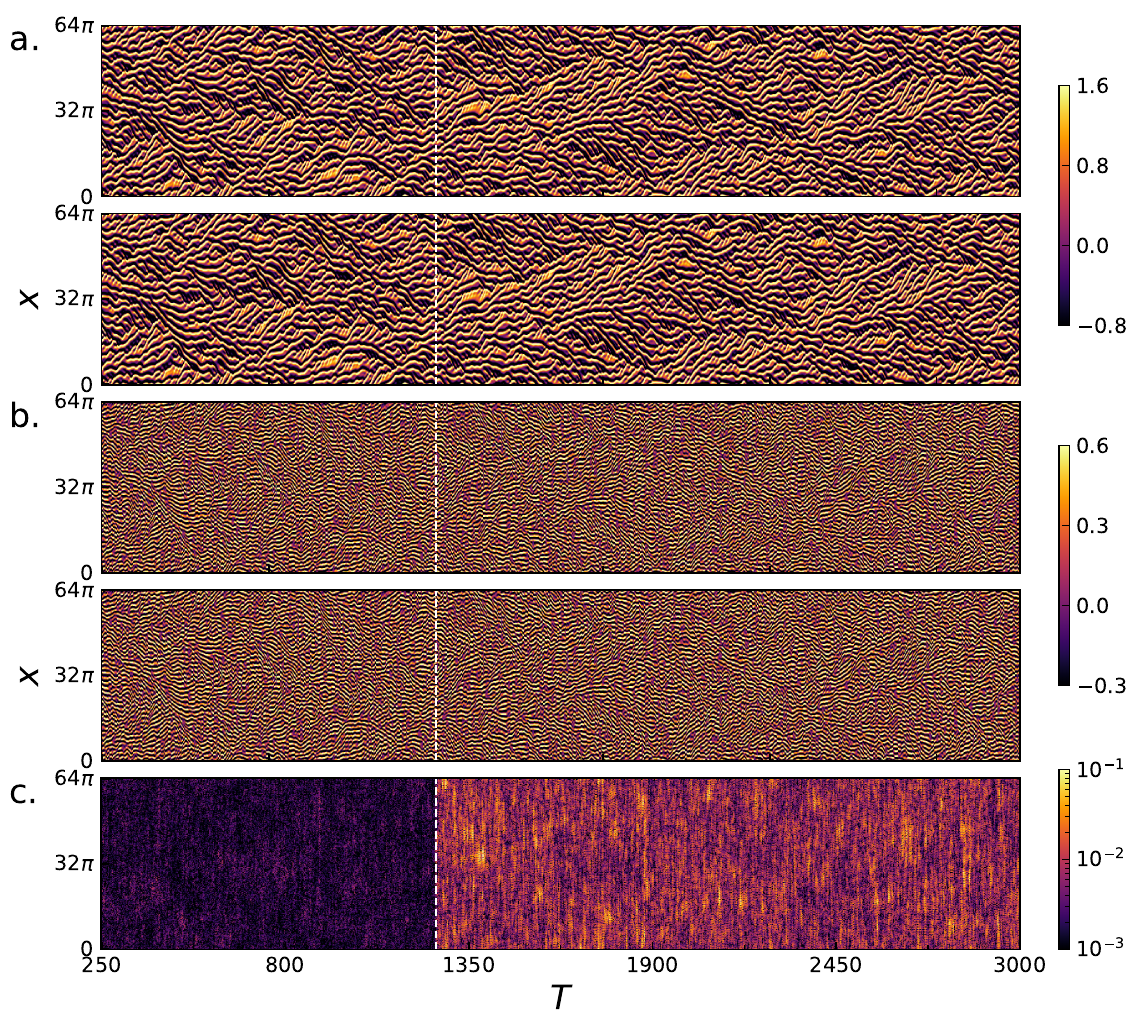}
  \caption{Illustration of dynamics of the Kuramoto-Sivashinsky equation in the full space $\mathscr{X}$ and complementary manifold $\mathscr{X}^{\perp}$. \textbf{a.} Real full dynamics (Top) and predicted full dynamics consist of real large-scale dynamics and predicted slaved small-scale dynamics (Bottom). \textbf{b.} Real slaved small-scale dynamics (Top) and predicted slaved small-scale dynamics (Bottom). \textbf{c.} Absolute error of small-scale slaved dynamics between truth and prediction. Left-hand side and right-hand side of vertical dash lines indicates corresponding quantities respectively on training dataset and test dataset. }
  \label{ks} 
\end{figure}

The performance of ND for solving the KS equation is comprehensively depicted in Fig. \ref{ks}. It is seen that, in Fig. \ref{ks}b, the ND reproduced the chaotic dynamics on $\mathscr{X}^{M^{\perp}}$ by solely large-scale dynamic on $\mathscr{X}^M$ with remarkable precision where absolute errors about $0.0018$ ($0.47\%$ in the relative $L_2$ errors) on training dataset and $0.0090$  ($2.44\%$ in the relative $L_2$ errors)  on test dataset are achieved by averaging them on the whole corresponding spatiotemporal domain. 
This also proves in an ML-assisted way that the small-scale dynamics are slaved really by the large-scale dynamics and can be represented precisely by a large-scale-dependent deterministic model to quantify the evolution of dissipative systems, in contrast to a stochastic mode\cite{schmuch2013}. 
It is observed that over a long-time horizon on the test dataset, the predicted results exhibit exceptional performance, showcasing a robust reproduction of the true small-scale nonlinear dynamics. This is evident not only in terms of alignment with dynamics in the physics space but also in capturing the true statistical properties of physics, including the power spectrum, autocorrelation, the second moment, and probability density function (refer to Section S4 of Supplementary information for details). Further investigation into the impact of the scale decomposition parameter $M$ is detailed in Section S5 of Supplementary information, hinting that the optimal complexity reduction, and the temporal-mixing step $k$ in Section S7 of Supplementary information, underscoring the crucial importance of considering temporal dependence of the small-scale dynamics.

\subsection{Forced Kuramoto-Sivashinsky equation}
In practical problems, external influences and perturbations usually exist, which can be taken as an external force acting on the system.
Subsequently, we investigate the following forced KS equation:
\begin{equation}\label{eq_fks}
\begin{aligned}
v_t+v_{x x}+v_{x x x x}+ vv_x=f,& \quad  \text{in} \  \mathcal{P} \times [0,T] \ ,\\
v(x, 0)=v_0(x),& \quad  \text{in} \  \mathcal{P} \ ,
\end{aligned}
\end{equation}
where $f(x)=\mu\cos(2\pi x/\lambda)$ is a forcing term introduced to break the symmetry\cite{pathak2018}, $\mathcal{P}:=]0,L[$ indicates the periodic domain, and $v_0(x)$ the initial condition. In the same way, pseudospectral method combined with the exponential time-differencing fourth-order Runge–Kutta formula is adopted to generate an $N$-step temporal sequence $\{v_i\}_{i=1}^{N}$. We now obtain numerical results with the initial condition $v_0=0$ and the chaos is alone determined by the forcing term. Same parameters $L = 64 \pi$, $T= 3000$ and $1024$ Fourier modes are employed, and $N=9000$ continuous snapshots are fetched starting from $750s$ when the chaos is fully developed. Then these snapshots are further split into the training dataset with the first $4000$ samples and the test dataset with the rest $5000$ samples. Through the same scale decomposition $M=33$, the continuous previous $t=5$ snapshots $\{\breve{v}_j\}^{i-1}_{j=i-4}$ and the current snapshot $\breve{v}_i$ together form a temporal sequence $\{\breve{v}_j\}^i_{j=i-4}$ as the input, and $\breve{v}_i^{\perp}$ as the target. 

\begin{figure}[htbp]
  \centering
  \includegraphics[width=0.8\textwidth]{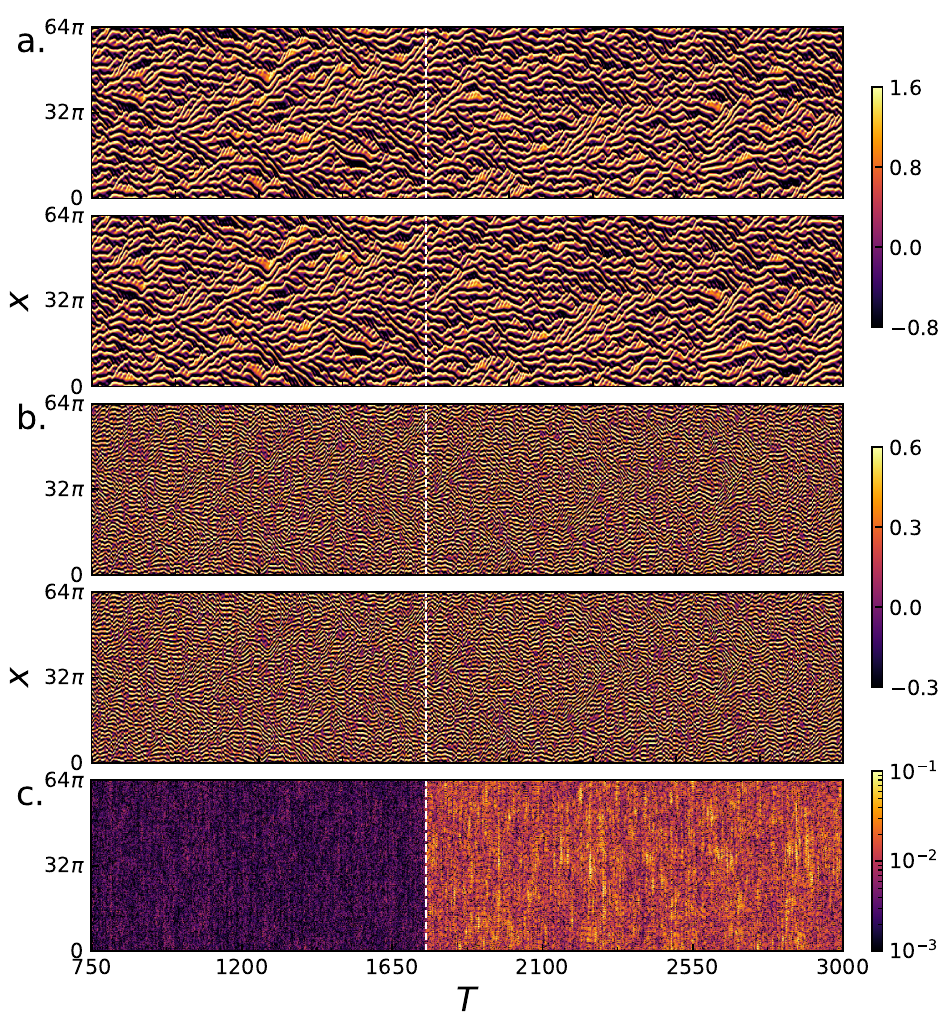}
  \caption{Illustration of dynamics of the forced Kuramoto-Sivashinsky equation in the full space $\mathscr{X}$ and complementary manifold $\mathscr{X}^{\perp}$. \textbf{a.} Real full dynamics (Top) and predicted full dynamics consist of real large-scale dynamics and predicted slaved small-scale dynamics (Bottom). \textbf{b.} Real slaved small-scale dynamics (Top) and predicted slaved small-scale dynamics (Bottom). \textbf{c.} Absolute error of small-scale slaved dynamics between truth and prediction. Left hand side and right hand side of vertical lines indicates corresponding quantities respectively on training dataset and test dataset. }
  \label{fks}
\end{figure}

The performance of ND for solving the forced KS equation is comprehensively depicted in Fig. \ref{fks}.
ND demonstrates impressive accuracy, similar to the unforced case, in reconstructing small-scale dynamics. It achieves absolute errors of approximately $0.0027$  ($0.69\%$ in the relative $L_2$ errors) on the training dataset and $0.0146$  ($3.79\%$ in the relative $L_2$ errors) on the test dataset when averaged over the entire corresponding spatiotemporal domain. Moreover, these promising results on a long time horizon in the test dataset prove that the existence of the inertial manifold for KS systems that completely describes the long-time dynamics, such that the neural operator can recognize the data distribution in the inertial manifold and learn the representation of small-scale dynamics by the inertial manifold. The approximation of the graph mapping (the inertial manifold) from a low-dimensional manifold to a complementary manifold is constructed successfully by the machine learning method, following the accurate reconstruction of the statistical properties of the power spectrum, autocorrelation, the second moment, and probability density function (refer to Section S4 of Supplementary information for details). The common conclusions on the insight into complexity reduction and the necessity of temporal-mixing steps are conducted in Section S5-S7 of Supplementary information.

\subsection{Forced 2D Navier-Stokes equations}
The Navier-Stokes equations have played a central role in enhancing our comprehension of fluid dynamics. They find application across a diverse range of scientific and engineering domains, including forecasting weather phenomena, modeling ocean currents, aircraft design, and analyzing blood flow dynamics. Within the framework of the Navier-Stokes system, highly chaotic dynamics and complex patterns emerge, exhibiting intricate spatial and temporal scales. As a more refined and illustrative demonstration of the comprehensive capabilities of the Navier-Stokes theory, particular attention is given to the following 2D equation in the vorticity formulation:
\begin{equation}
\begin{aligned}
\partial_t w(x, t)+u(x, t) \cdot \nabla w(x, t) & =\nu \Delta w(x, t)+f(x), & & \text{in} \ \mathcal{T^2} \times [0,T] \\
\nabla \cdot u(x, t) & =0, & & \text{in} \ \mathcal{T^2} \times [0,T] \\
w(x, 0) & =w_0(x), & & \text{in} \ \mathcal{T^2},
\end{aligned}
\end{equation}
where $u$ represents the velocity, $w=\nabla \times u$ represents the vorticity, $\nu$ represents the viscosity coefficient and $\mathcal{T^2}:= \mathcal{P} \times \mathcal{P}:= ]0,L[ \times ]0,L[$ indicates the 2-torus defined as the product of two periodic physics domain, and $w_0(x)$ the initial condition. Similarly, pseudospectral method coupled with the exponential time-differencing fourth-order Runge–Kutta formula is adopted to generate an $N$-step temporal sequence $\{v_i\}_{i=1}^{N}$. Assuming $\hat{f}$ and $\hat{w}$ are the Fourier transform for $f$ and $w$, respectively, and then $f(x)$ is set to be $\hat{f}(k)=\frac{\hat{w}(k,t)}{20}, \ 4\leq k \leq5$, where k indicates the wave vector. The vorticity is driven by itself truncated at a specified range. Numerical simulation is conducted with parameters of $\nu=1e-3$, $L=2\pi$, $T=85000$ and $64$ Fourier modes, and the random initial condition is set to $w_0(x)=1+2\cdot\mathcal{U}((0,1)^2)$ where $\mathcal{U}$ represents the uniform distribution. To capture the latent quasi-periodic/periodic dynamics underlying the forced NS system, a total of $N=8000$ snapshots ranging from $T=5000$ to $T=85000$ are generated, and the initial $4000$ snapshots are split to train the neural operator and the left 4000 snapshots are split to form the test dataset. Through scale decomposition with $M=13$, the continuous previous $t=4$ snapshots $\{\breve{v}_j\}^{i-1}_{j=i-4}$ and the current snapshot $\breve{v}_i$ together form a temporal sequence $\{\breve{v}_j\}^i_{j=i-4}$ as the input, and $\breve{v}_i^{\perp}$ as the target  of neural network.

\begin{figure}[htbp]
  \centering
  \includegraphics[width=0.95\textwidth]{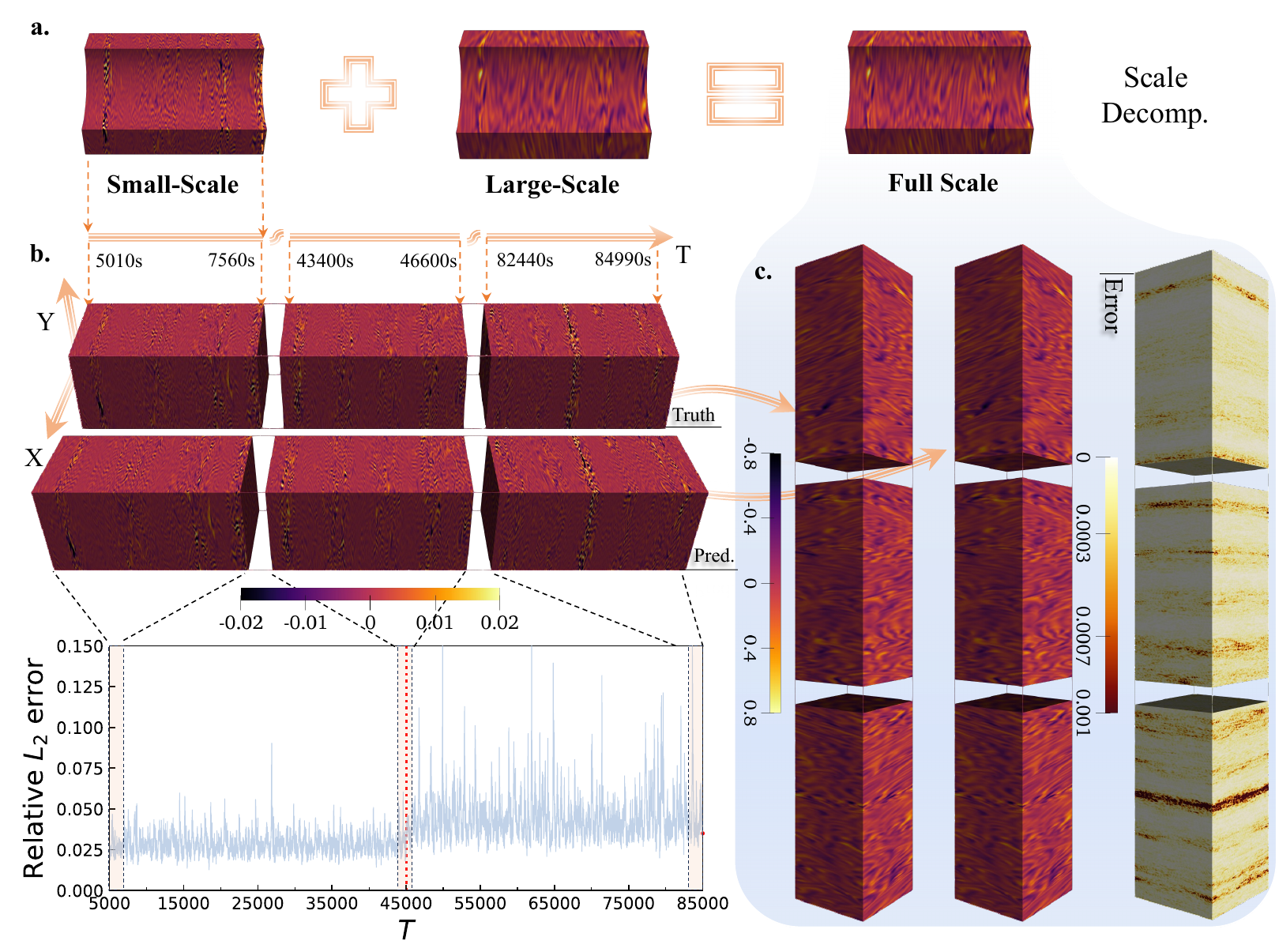}
  \caption{Illustration of forced 2d Navier-Stokes equation with $\nu=1e-3$. \textbf{a}. Illustration of the scale decomposition of the starting temporal sequence in $[5010, 7560]$. \textbf{b}. Real and predicted small-scale dynamics in three different time intervals and relative $L_2$ error averaged on time cross-section where the left-hand side and right-hand side of the red vertical line indicates corresponding quantities respectively on training dataset and test dataset. \textbf{c}. Comparison between real and predicted full-scale dynamics in three different time intervals.}
  \label{forced_turb2d_t}
\end{figure}

\begin{figure}[htbp]
  \centering
  \includegraphics[width=0.95\textwidth]{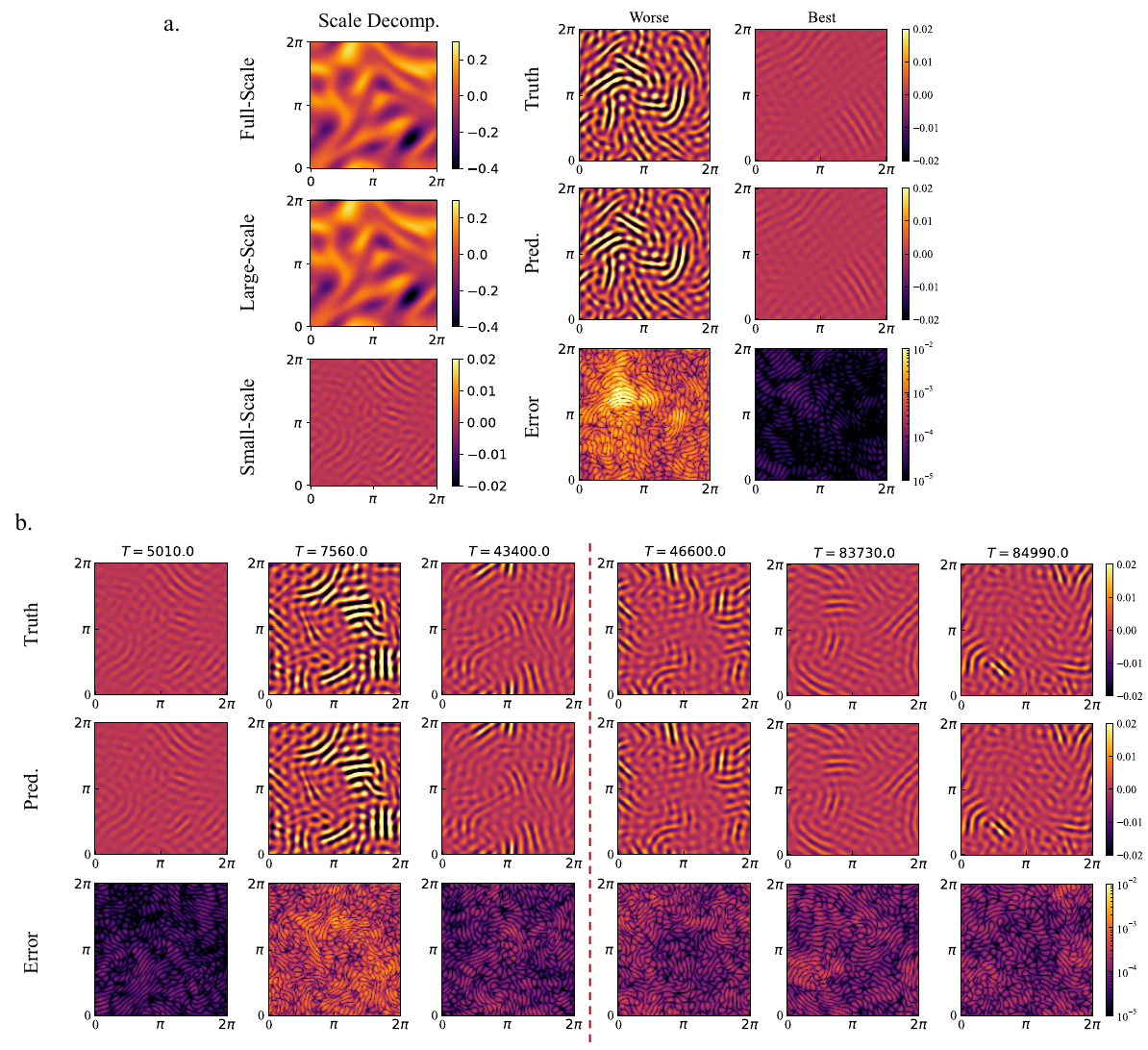}
  \caption{Illustration of dynamics in the physical domain. a. The scale decomposition at $T=5010$ and two time slices representing extremes in the precision of the ND's generalization on test dataset are shown. b. A comparison of slaved small-scale dynamics between truth and prediction at truncated ends of partial temporal sequences is shown with an absolute error, where the left-hand side and right-hand side of the red vertical line indicates corresponding quantities respectively on training dataset and test dataset.}
  \label{forced_turb2d}
\end{figure}

The performance of ND for solving the forced 2d NS equation is comprehensively illustrated in Fig.\ref{forced_turb2d_t} and Fig.\ref{forced_turb2d}. As observed in Fig. \ref{forced_turb2d_t}b, the first time step comes to $5010$ instead of $5000$, which is caused by the assembly of previous time steps for time-dependent input, and the absolute error averaged on the spatiotemporal domain achieves approximately $9.88e-5$ ($2.83\%$ in the relative $L_2$ errors) and $1.64e-4$ ($4.47\%$ in the relative $L_2$ errors) on the training dataset and test dataset, respectively. It is worth mentioning that the steady prediction precision on a broad time range is triggered again with the excellent reconstruction of statistical properties (refer to Section S4 of Supplementary information for details). Even in such a more sophisticated system, it preserves the conclusion that temporal-mixing steps are necessary and sheds light on complexity reduction as shown in Section S5-S7 of Supplementary information.

Concentrating on the dynamics in the physics space, the original vorticity and scale decomposition counterparts at the first time step $T=5010$ are diagrammed in Fig. \ref{forced_turb2d}a, following comparison for the test dataset between the moments, where one is the worse precision with about $1.02e-3$ averaged absolute error ($16.24\%$ relative $L_2$ error) and another one is the best precision with about $1.70e-5$ averaged absolute error ($1.99\%$ relative $L_2$ error). In Fig. \ref{forced_turb2d}b, the precisions at the truncated time steps, which are aligned with the ends of each partial temporal sequence, are also included, showcasing ND's prediction capability on training dataset (right-hand side of the red vertical dash line) and test dataset (left-hand side of the red vertical dash line). It can be observed that quasi-periodic dynamics emerge companying a periodic intensity adjustment and the absolute error changes with the change of tiny structure intensity. However, even in the worse slice, the absolute error remains at a very low order, peaking at a mere $1e-3$ at its maximum. Ultimately, the results underscore that, when dealing with the presence of long-time dynamics embedded in the dynamics system, ND would recognize the low-dimensional manifold (embedding) composed of the trajectories of large-scale dynamics and approximate the inertial manifold with a promising performance in capturing the intricate small-scale dynamics. Moreover, the study on the free-decaying NS system is also performed to investigate the ND's prediction capability on the complex system characterized by the short-term dynamics in Section S1 of Supplementary information.

\section{Discussion}\label{sec12}
The effectiveness of the ND has been demonstrated in tackling fluid dynamics problems with uniform meshes and regular geometries. While these cases serve as foundational examples, our method's adaptability extends beyond such straightforward scenarios.
In subsequent discussions, we will explore the potential application of our approach to more complex problems.

A common concern often arises regarding the handling of \textbf{non-uniform meshes}. One approach is to devise a mesh mapping operator, akin to the encoder structure, to transform non-uniform meshes into uniform ones. Conversely, an inverse mesh mapping operator (decoder operator) can be developed to revert uniform meshes back to non-uniform ones if the prediction mesh format needs to align with the input format. 
To facilitate training and utilization, the mesh mapping operators and the ND can be integrated into a cohesive operator structure. 
Consequently, the input for the ND transitions to the latent representation of non-uniform meshes generated by the encoder operator, while the output shifts to the latent representation of uniform meshes. This output is then fed into the decoder operator to transform it back to non-uniform meshes. A similar approach has been successfully implemented and is detailed in \cite{li2023fourier}.

Addressing contaminated data stemming from \textbf{noisy large-scale dynamics} stands as a paramount challenge within the machine learning community. In this work, we propose several methods to mitigate this issue. Employing pre-processing techniques for contaminated data emerges as a favorable approach. By leveraging repetitive mode decomposition procedures, such as spectral decomposition, it becomes feasible to extract the original dynamics and attenuate high-frequency noise effectively. In scenarios where low-frequency noise predominates, regularization methods such as dropout\cite{srivastava2014dropout} and batch normalization\cite{ioffe2015batch} can enhance generalization performance. 
Alternatively, a noise adaptation layer can be integrated atop the ND architecture to learn the noise transition process, while developing a dedicated architecture capable of reliably accommodating a broader spectrum of data noise types.

ND holds significant promise in addressing the concept of \textbf{causal emergence} through facilitating the scale transformation from small-scale to large-scale dynamics. While our current focus may be on realizing the scale transformation from large-scale to small-scale, the inherent capabilities of our approach make it well-suited for tackling the inverse problem as well. By using historical information and considering temporal-dependent input data, our method grasps the intricate relationship between small-scale and large-scale dynamics. This allows us to capture how collective behaviors at the large scale emerge from microscopic interactions. Actually, previous studies had been performed to investigate such a scale transformation, for example, the equation-free framework\cite{kevrekidis2003equation}, the heterogeneous multiscale method\cite{weinan2007heterogeneous} and ML-assisted effective dynamics learning\cite{vlachas2022multiscale}. These works bring us confidence that developing an inverse version of ND will be possible.

\section{Methods}\label{Methods}
\subsection*{Operator form of partial differential equations}\label{Operator form of partial differential equations}
We consider the follow general classical form of partial differential equations (PDEs) in an operator form in a function space $\mathscr{X}$ with a finite time horizon $T\in \mathbb{R}^+$:
\begin{equation}\label{eq:op}
\left\{\begin{array}{rll}
\displaystyle \frac{d u}{d t}&=\mathscr{A}(u),  & \text{in}\ T\times D\in\mathbb{R}^d \\
u&=g,& \text{in}\  T\times \partial D \\
u&=u_0,& \text{in}\ \{0\}\times D\in\mathbb{R}^d
\end{array}
\right.
\end{equation}
where $u: T\times D\rightarrow \mathscr{X}$ is the solution,  $\mathscr{A}$ is a nonlinear partial differential operator with domain $\mathscr{D}(\mathscr{\mathscr{X}})\in \mathscr{X}$ , $D$ is a bounded domain in $\mathbb{R}^d$ ($d\in \mathbb{Z}^+$ denotes the spatial dimension), and $u_0$ and $g$ denote respectively the initial and boundary condition.  For simplicity, control parameters are not introduced explicitly in Eq.\ref{eq:op}. Here, $\mathscr{X}$ is generally a Banach space equipped with norm $\|\cdot\|_{\mathscr{X}}$.  In practice, $\mathscr{X}$ is equipped with the $L^2$ norm as $\|\cdot\|_{L^2(\mathscr{X})}$.

Classically,  a solution operator $\mathscr{G}^t$ of PDEs can be defined as $\mathscr{G}^tu(0)=u(t)$.  If $T$ is evenly partitioned as $\tau=\{t_k\}_{k=0}^K$ with a time interval $\delta t$,  without loss of generality,  we define $\mathscr{G}^{\delta t} $ as a discrete solution operator such that  $\mathscr{G}^{\delta t} u_{i}=u_{i+1}$ where $u_i$ denotes $u(t_i)$. In semi-group sense, it is straightforward that $\mathscr{G}^su(t)=\mathscr{G}^s\circ\mathscr{G}^tu(0)=u(t+s)=\mathscr{G}^{t+s}u(0)$. Considering the integral form,  $\mathscr{G}^t$ can be defined classically as follows:
\begin{equation}
    \mathscr{G}^tu(0)=u_0+\int_{0}^{t}\mathscr{A}(u)ds.
\end{equation}

\subsection{Low-dimensional models}\label{Low-dimensional models}
When the evolution system of Eq. \ref{eq:op} is studied experimentally or numerically,  the obtained data usually lies in a finite-dimensional space. Even though the dynamic system itself is {\it a priori} infinite-dimensional, there is a belief that it possesses a global attractor in a low-dimensional space. This belief stems from rigorous demonstrations in several partial differential equations (PDEs) showing that the dynamics exponentially converge onto a global attractor within a low-dimensional subset of the phase space\cite{foias1988inertial,temam2012infinite}. Therefore,  such systems do possess low-dimensional dynamics,  particularly for the Navier-Stokes equations (NSEs).

Supposing that for the Eq. \ref{eq:op}, there exists an orthogonal basis space $\mathscr{X}^\infty=\{\phi_i\}_{i=1}^\infty$,  then the following modal decomposition of $u$ can be introduced:
\begin{equation}
    u=\sum_{i=1}^\infty a_i\phi_i,
\end{equation}
where $a_i$ is theoretically determined by the inner product $\langle u,\phi_i\rangle_{\mathscr{X}^\infty}$ in $\mathscr{X}^\infty$. If the nonlinear evolution system \ref{eq:op} has a smooth, invariant, inertial manifold $\mathscr{M}$,  it may be parameterised by a finite-dimensional subspace represented only by the first $M$ modal coefficients $\{a_i\}_{i=1}^M$. Hereby,  a reduction approach with a finite set of $\{a_i\}_{i=1}^M$ determines the finite-dimensional dynamic property.  We here define $\mathscr{P}_M$ as an $L^2$-orthogonal projection operator from $\mathscr{X}^\infty$ to $\mathscr{X}^M$.  Then $\mathscr{P}_Mu$ denotes a representation of $u$ in a finite-dimensional space $\mathscr{X}^M$,  which contains most important dynamic information of the original one to describe the long-time dynamics assuming $\mathscr{X}^M$ is chosen properly.

\subsection{Scale decomposition on dynamics}\label{Scale decomposition on dynamics}
Practically, $\mathscr{X}^M$ can be constructed through Fourier modes decomposition, proper orthogonal decomposition (POD) eigenmodes\cite{Lumley1967, holmes2012turbulence}, Koopman modes\cite{mauroy2020koopman}, operator normal modes\cite{vakakis2001normal},  orthogonal function spaces or various mode decomposition methods\cite{rowley2017model}. In our study, the Fourier space is employed to conduct the scale decomposition through the differentiation of fast modes, which are characterized by large wave numbers that rapidly converge towards equilibrium in contrast to the slow (low wave number) modes\cite{schmuch2013}.  Given the rapidity and convenience afforded by fast Fourier transformation, Fourier modes decomposition is adapted to separate the scale of original (full-scale) data, and the corresponding basis functions form the low-dimensional orthogonal space and the complementary space, respectively.

Utilizing the Fourier decomposition method, scale decomposition on dynamics can be achieved by truncating a specified number of Fourier modes. Assuming $\{u_i\}_{i=1}^{N}$ is the temporal snapshot dataset of a dynamic system,  each $u_i$ can be represented exactly in a Fourier mode space $\mathscr{X}$=$\{\exp(-\imath j x) \}_{j\in\mathbb{Z},\  |j|\leq{N_d}}$, where ${N_d}\in \mathbb{N^+}$ is the finite space dimension and depends on the resolution of dataset since $u_i$ is approximated numerically in a set with finite Fourier modes. Correspondingly, $\mathscr{P}_M$ turns to an $L^2$-orthogonal projection operator from $\mathscr{X}^{N_d}$ to $\mathscr{X}^M$ in the numerical context. Then $u_i$ in $\mathscr{X}$ can be represented by:
\begin{equation}
    u_i=\sum_{|j|\leq{N_d}}\hat{u}_{i,j}\exp(-\imath j x).
\end{equation} 
Naturally, the subspace $\mathscr{X}^{M}$=$\{\exp(-\imath j x) \}_{|j|\le M}$ and $\mathscr{X}^{M^{\perp}}$=$\{\exp(-\imath j x) \}_{M<|j|\le {N_d}}$ can be further defined where ${M}$ is finite, the $u_i$ in $\mathscr{X}^{M}$ and $\mathscr{X}^{M^{\perp}}$ are in the form of 
\begin{equation}
    \mathscr{P}_Mu_i=\sum_{|j|\leq M}\hat{u}_{i,j}\exp(-\imath j x).
\end{equation} 
and
\begin{equation}
    \mathscr{Q}_Mu_i=\sum_{M<|j|\leq {N_d}}\hat{u}_{i,j}\exp(-\imath j x).
\end{equation} 
Consequently, the large-scale dynamics $\mathscr{P}_Mu_i$ and small-scale dynamics $\mathscr{Q}_Mu_i$ are captured, resulting in the temporal snapshot dataset of large-scale dynamics $\{\mathscr{P}_Mu_i\}_{i=1}^{N}$ and small-scale dynamics $\{\mathscr{Q}_Mu_i\}_{i=1}^{N}$.

\subsection{Neural operator}\label{Neural operator}
Considering the spatial region $D \subset \mathbb{R}^d$ (with $d$ being the dimension of the space), an operator mapping from parameter space to solution space is defined as follows\cite{kovachki2023neural}:
\begin{equation}
    \mathcal{G}:\mathcal{A}(D;\mathbb{R}^{d_a}) \rightarrow \mathcal{U}(D;\mathbb{R}^{d_u})
\end{equation}
where $\mathcal{A}(D;\mathbb{R}^{d_a})$ and $\mathcal{U}(D;\mathbb{R}^{d_u})$ are Banach spaces and $d_a, d_u \in \mathbb{N}$ are the corresponding dimension. For $a \in \mathcal{A}(D;\mathbb{R}^{d_a})$ and $u\in \mathcal{U}(D;\mathbb{R}^{d_u})$, it has the relation $a \rightarrow u := \mathcal{G}(a)$.
Then, the mapping form of the neural operator can be defined as follows:
 \begin{equation}
 \begin{aligned}
     \mathcal{G_{\rho}}:\mathcal{A}(D;\mathbb{R}^{d_a}) &\rightarrow \mathcal{U}(D;\mathbb{R}^{d_u}), \\
     a &\rightarrow \mathcal{G_{\rho}}(a).
 \end{aligned}
 \end{equation}
Given a depth $L\in \mathbb{N}$ for neural operator, the neural operator takes the following form:
\begin{equation}
    \mathcal{G_{\rho}}(a) = \mathcal{Q} \circ \mathcal{L}_L \circ \mathcal{L}_{L-1} \circ \cdots \circ \mathcal{L}_1 \circ \mathcal{R}(a)
\end{equation}
where $\mathcal{R}:\mathcal{A}(D;\mathbb{R}^{d_a}) \rightarrow \mathcal{U}(D;\mathbb{R}^{d_v}) \ (d_a \leq d_v)$ indicates the lifting operator in the form $\mathcal{R}(a)(x)=Ra(x), \ R \in \mathbb{R}^{d_v \times d_a}$. $\mathcal{Q}:\mathcal{U}(D;\mathbb{R}^{d_v}) \rightarrow \mathcal{U}(D;\mathbb{R}^{d_u})$ indicates the local projection operator in the form $\mathcal{Q}(v)(x)=Qv(x), \ Q \in \mathbb{R}^{d_u \times d_v}.$
Similar to general finite-dimensional neural networks, the nonlinear operator $\mathcal{L}_\mathscr{l}:\mathcal{U}(D;\mathbb{R}^{d_v}) \rightarrow \mathcal{U}(D;\mathbb{R}^{d_v})$ is defined as 
$ \mathcal{L}_{\mathscr{l}}(v)(x)=\sigma(W_{\mathscr{l}} v(x) + b_{\mathscr{l}}(x) + (\mathcal{K}(a;\theta_{\mathscr{l}})v)(x), \ \forall x \in D,$
where the nonlinear activation function $\sigma$ is defined component-wise, the weight matrix $W_{\mathscr{l}} \in \mathbb{R}^{d_v \times d_v}$ and bias $b_{\mathscr{l}} \in \mathcal{U}(D;\mathbb{R}^{d_v})$ are used to define the affine pointwise mapping as $W_{\mathscr{l}} v(x) + b_{\mathscr{l}}(x)$.

The global linear operator $\mathcal{K}:\mathcal{A} \times \Theta \rightarrow L(\mathcal{U}(D;\mathbb{R}^{d_v}), \mathcal{U}(D;\mathbb{R}^{d_v}))$ maps the input field $a$ and parameters $\theta \in \Theta$ to the linear operator $\kappa(a;\theta):\mathcal{U}(D;\mathbb{R}^{d_v}) \rightarrow \mathcal{U}(D;\mathbb{R}^{d_v})$. The classical form of the integral operator  is given by:
\begin{equation}
    (\mathcal{K}(a;\theta)v)(x)=\int_{D} \kappa_{\theta}(x,y,a(x),a(y))v(y) dy, \  \forall x \in D,
\end{equation}
where integral kernel function $\kappa_{\theta}: \mathbb{R}^{2(d+d_a)} \rightarrow \mathbb{R}^{d_v \times d_v}$ is a neural network parameterized by $\theta \in \Theta$. Consequently, the concept of neural operator is clarified and a definition is provided, and then $\mathcal{L}_{\mathscr{l}}$ refers to as neural layer and $\mathcal{G_{\rho}}$ neural operator. More details about neural operator structure and parameters adopted for numerical examples can refer to Section S2 of Supplementary information.

\backmatter
\section{Conclusions}
To enhance the efficiency and precision of the multiscale simulation, the well-represented large dynamics are leveraged to assist the modelling of small-scale dynamics. A pioneering methodology termed Neural Downscaling (ND) is proposed based on the theory of inertia manifold. ND establishs a neural-operator-assisted graph mapping from large-scale (slow) dynamics that locate at a low dimensional manifold, to slaved small-scale (fast) dynamics that locate at a complementary manifold.
In contrast to implicit and stochastic modeling approaches typically employed for arbitrary small-scale dynamics, and explicit and deterministic modeling methods primarily used for the Navier-Stokes (NS) system, ND offers the capability to construct an explicit and deterministic model that accommodates a wide range of small-scale dynamics.

Numerical investigations demonstrated the validity and high fidelity of ND in nonlinear spatiotemporal complex systems, as governed by the Kuramoto-Sivashinsky (KS) and Navier-Stokes (NS) equations. These reconstructions of small-scale dynamics not only adhere to the principles of statistical physics but also exhibit remarkable generalization capabilities. Furthermore, a comprehensive investigation of the scale decomposition offers new perspectives on the interrelationships between large-scale and small-scale dynamics and sheds light on the complexity reduction to guide the omission of small-scale dynamics. It is reasonable to believe that ND is crucial for understanding and modelling physical process of complex systems. It holds promise for addressing multi-scale problems and extending to diverse fields for similar problems such as causal emergence, thereby pushing the boundaries of our comprehension of complex spatiotemporal nonlinear systems and offering a potential breakthrough in achieving a deeper understanding of intricate system behaviors.

\bmhead{Supplementary information} Forced 2D Navier-Stokes equations.mp4, Free-decaying 2D Navier-Stokes equations.mp4, Supplementary information.pdf. This PDF file includes Sections S1 to S7, Table 1 to 2, Fig. 1 to 8, References 2.

\bmhead{Acknowledgments}

This work is supported by the Natural Science Foundation of China (Nos. 91630205, 91852106, 92152301).

\section*{Declarations}

The authors declare no conflict of interest.


\bibliography{sn-bibliography}


\end{document}


\title[Supplementary Information ]{Supplementary information 

for

Neural Downscaling for Complex Systems: from Large-scale to Small-scale by Neural Operator}


\author[1]{\fnm{Pengyu} \sur{Lai}}\email{laipengyu@sjtu.edu.cn}

\author[1]{\fnm{Jing} \sur{Wang}}

\author[1]{\fnm{Rui} \sur{Wang}}

\author[1]{\fnm{Dewu} \sur{Yang}}

\author[1]{\fnm{Haoqi} \sur{Fei}}

\author*[1]{\fnm{Hui} \sur{Xu}}\email{dr.hxu@sjtu.edu.cn}

\affil*[1]{\orgdiv{School of Aeronautics and Astronautics}, \orgname{Shanghai Jiao Tong University}, \orgaddress{\city{Shanghai}, \postcode{200240}, \country{China}}}

\maketitle
\newpage
\section{Section S1. Free-decaying 2D Navier-Stokes equations}
To investigate the ND's performance on the short-time dynamics, the free-decaying 2D Navier-Stokes (NS) equations for a viscous, incompressible fluid in vorticity form is presented as follows:
\begin{equation}
\begin{aligned}
\partial_t w(x, t)+u(x, t) \cdot \nabla w(x, t) & =\nu \Delta w(x, t), & & \text{in} \ \mathcal{T^2} \times [0,T] \\
\nabla \cdot u(x, t) & =0, & & \text{in} \ \mathcal{T^2} \times [0,T] \\
w(x, 0) & =w_0(x), & & \text{in} \ \mathcal{T^2}
\end{aligned}
\end{equation}
where notations are consistent with those employed in the forced NS equation. Similarly, pseudospectral method coupled with the exponential time-differencing fourth-order Runge–Kutta formula is adopted to generate an $N$-step temporal sequence $\{v_i\}_{i=1}^{N}$. Then the numerical calculation is performed with parameters of $\nu=5e-4$, $L=2\pi$, $T=70$, and $64$ Fourier modes, and the same random initial is utilized as adopted in the forced NS equation. We are interested in the time horizon that encompasses moments deviating from the random initial state, showcasing distinctive vorticity structures with appropriate energy. Consequently, a total of $N=2000$ snapshots are gathered over the time from $50s$ to $70s$, and the initial $1600$ snapshots are selected to construct the training dataset and the last $400$ snapshots are for test dataset. Through scale decomposition with $M=5$, the continuous previous $t=4$ snapshots $\{\breve{v}_j\}^{i-1}_{j=i-4}$ and the current snapshot $\breve{v}_i$ together form a temporal sequence $\{\breve{v}_j\}^i_{j=i-4}$ as the input, and $\breve{v}_i^{\perp}$ as the target  of neural network. 

\begin{figure}[htbp]
  \centering
  \includegraphics[width=0.95\textwidth]{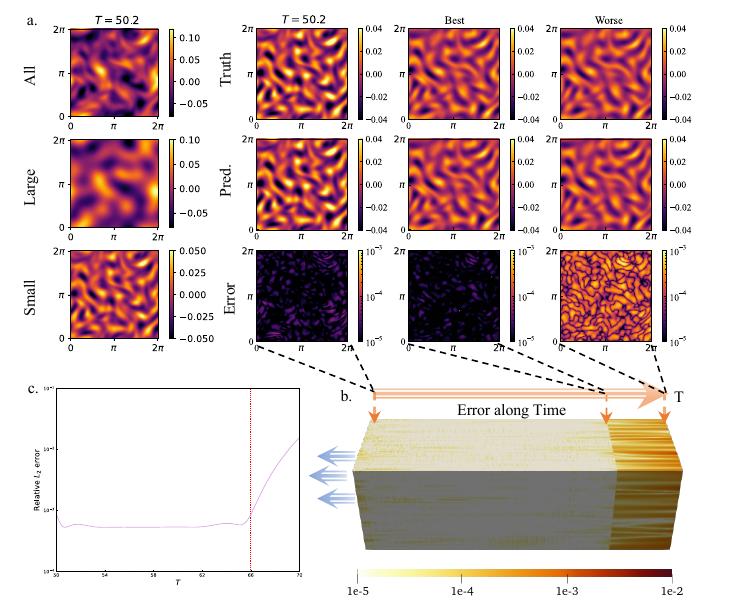}
  \caption{Illustration of free-decaying 2d Navier-Stokes equation with $\nu=5e-4$. \textbf{a}. Illustration of the scale decomposition at $T=50$ in the first column. Comparison of slaved small-scale dynamics between truth and prediction at $T=50$, $T=66$ and $T=70$ is shown with an absolute error, where $T=50$ is located on the training dataset and $T=66$ and $T=70$ on the test dataset. The two selected time slices from the test dataset represent extremes in the precision of ND's generalization. \textbf{b}. Absolute error of slaved small-scale dynamics in the time interval $[50, 70]$. \textbf{c}. Relative $L_2$ error averaged on time cross-section. Left-hand side and right-hand side of vertical lines indicates corresponding quantities respectively on training dataset and test dataset.}
  \label{turb2d}
\end{figure}

In Fig. \ref{turb2d}a, the first column illustrates the original vorticity and the scale decomposition counterparts at $T=50.2$, and the subsequent three columns depict the absolute errors between ground truths and predictions at $T=50.2$, $T=66$ and $T=70$, respectively.  Notably, the last two predictions correspond to the test dataset, with $T=66$ representing the best performance of $8.84e-4$ in absolute error ( $0.09\%$ in relative $L_2$ error) and $T=70$ showcasing the worst performance of $1.48e-4$ in absolute error ( $1.56\%$ in relative $L_2$ error). To provide a comprehensive overview of ND's performance, Fig. \ref{turb2d}(b) illustrates the errors across the entire dataset, offering a time-evolutionary visualization. Additionally, Fig. \ref{turb2d} c displays the relative $L_2$ error averaged on the slice, highlighting ND's efficacy in the physics space. ND achieves approximately $6.79e-6$ in absolute error ($0.06\%$ in relative $L_2$ error) on training dataset and $5.82e-5$ in absolute error ($0.60\%$ in relative $L_2$ error) on test dataset. It is noteworthy that the error increases over time on test dataset, attributed to the dissipative characteristics of this unforced system. The dynamics governed by the Navier-Stokes equation, without the introduction of external energy, naturally decay over time as the system's energy cascades from broad large-scale fluctuations to small-scale ones, ultimately dissipating as heat. The short-time dynamics recognized by neural operators prove insufficient in describing the entire dissipative process based solely on limited data learning at $[50, 66]$, in contrast to the forced NS example where adequate data are obtained to capture the underlying long-time dynamics. If a dataset encompassing the entire dissipative process is generated for training ND, it is evident that dynamics predictions at any moment do not involve the time generalization beyond the training data distribution, and then the accurate prediction absolutely can be achieved, eliminating the challenge of extrapolating to out-of-distribution scenarios. Therefore, the presented results highlight the potential of ND, demonstrating that ND can generalize to unknown dynamics to a certain extent with a promising fidelity.

\newpage
\section{Section S2. Structure and parameters of neural operator}
Taking the Fourier neural operator (FNO) \cite{li2020fourier} as an instantiation of spectral neural operator in this paper, FNO is a special form of the spectral neural operator with a kernel function $\kappa_{\theta}(x,y,a(x),a(y))$ =$\kappa_{\theta}(x-y)$ . The integral form of the global linear operator $\mathcal{K}(a,\theta)$ can be written in the following convolution form:
 \begin{equation}
    (\mathcal{K}(a;\theta)v)(x)=\int_{D} \kappa_{\theta}(x-y)v(y) dy, \  \forall x \in D.
\end{equation}
Naturally, the convolution operation can be calculated by the Fourier transform as follows:
\begin{equation}
    (\mathcal{K}(a;\theta)v)(x)= \mathcal{F}^{-1} (P_{\theta}(k) \cdot \mathcal{F}(v)(k))(x), \ \forall x \in \mathbb{T}^d
\end{equation}
where $P_{\theta}(k) \in \mathbb{C}^{d_v \times d_v}$ is derived from the Fourier transform of the integral kernel function, i.e. $P_{\theta}(k)=\mathcal{F}(\kappa_{\theta})(k)$, and defined in a periodic domain $\mathbb{T}^d$. The nonlinear operator $\mathcal{L}_{\mathscr{l}}$ turns to the following form:
\begin{equation}
     \mathcal{L}_{\mathscr{l}}(v)(x)=\sigma(W_{\mathscr{l}} v(x) + b_{\mathscr{l}}(x) + \mathcal{F}^{-1} (P_{\mathscr{l}}(k) \cdot \mathcal{F}(v)(k))(x)), \ \forall x \in D.
 \end{equation}
In theory, FNO has the following universal approximation property\cite{chen1995universal}:
\begin{theorem}($\varepsilon$-approximation of neural operator) Suppose that $s,s'>0$, $\mathcal{G}:H^s(\mathbb{T}^d;\mathbb{R}^{d_a}) \rightarrow H^{s'}(\mathbb{T}^d;\mathbb{R}^{d_u})$ is a nonlinear continuous operator, $K \subset H^s(\mathbb{T}^d;\mathbb{R}^{d_a})$ is a compact support set. Then for any $\epsilon>0$, there exists an neural operator $G_{\rho}:H^s(\mathbb{T}^d;\mathbb{R}^{d_a}) \rightarrow H^{s'}(\mathbb{T}^d;\mathbb{R}^{d_u})$ such that
$$
\sup _{a \in K}\|\mathcal{G}(a)-\mathcal{G_\rho}(a)\|_{H^{s^{\prime}}} \leq \epsilon.
$$
\end{theorem}

\begin{figure}[htbp]
  \centering
  \includegraphics[width=0.95\textwidth]{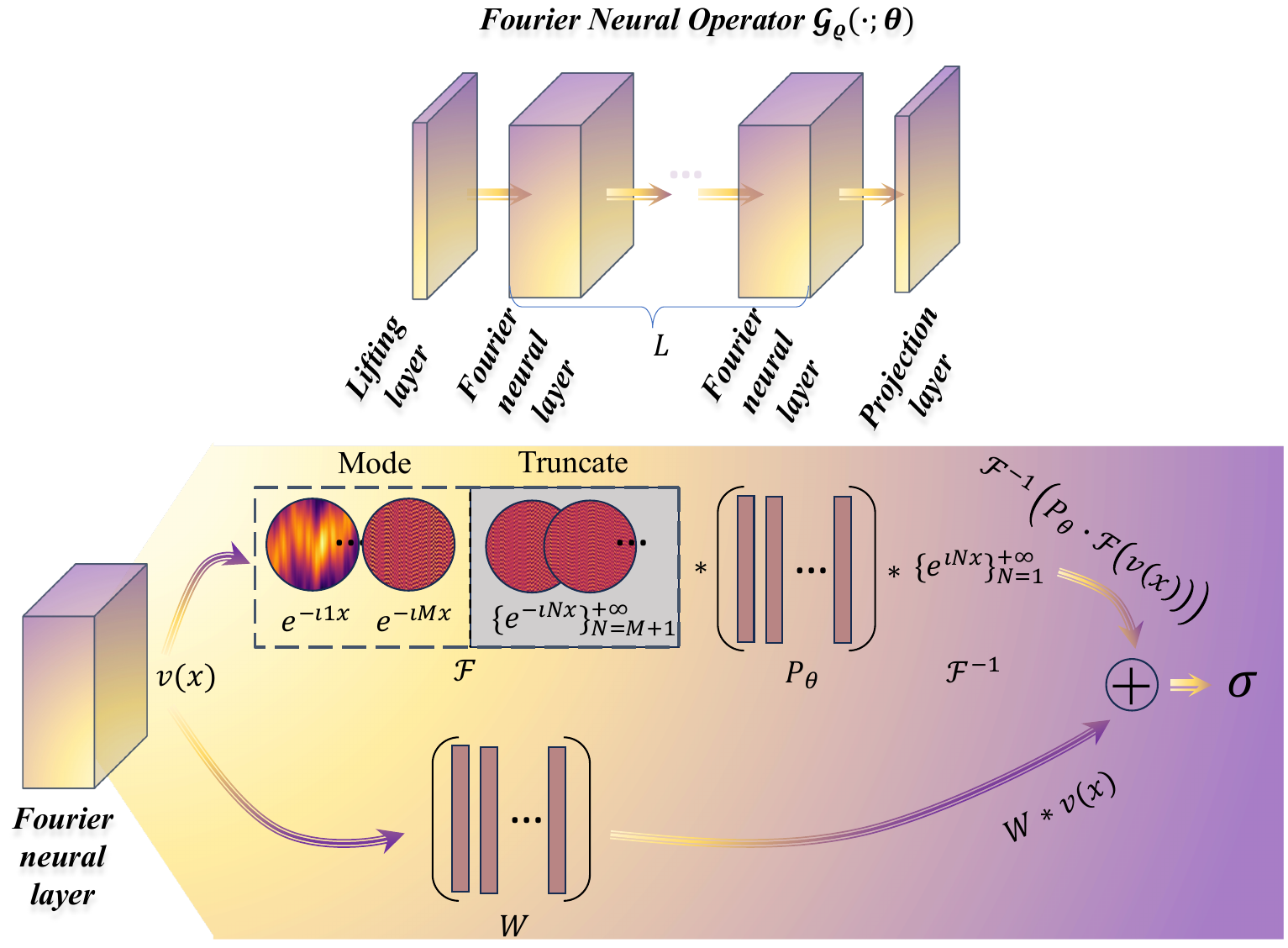}
  \caption{Illustration of Fourier neural operator.}
  \label{fno}
\end{figure}

The Fourier neural operator (FNO) is as shown in Fig.\ref{fno}. As mentioned above, it consists of a lifting layer, $L$ Fourier neural layers, and a projection layer, where the Fourier neural layer refers to the nonlinear operator $\mathcal{L}_{\mathscr{l}}$. Especially, the physical position, as the grid discretization information, is embedded into the channel in the lifting layer to promote the ND training.

The fully connected neural network is considered for the lifting layer in the KS equation and forced KS equation, which is used to lift the input (channel) dimension to the desired input (channel) dimension of the Fourier neural layer. The two-layer convolutional neural network (CNN) with 1 kernel size, called convolutional block (CB), is considered for the lifting layer in the free-decaying NS equations and the forced NS equations, owing to the two-dimensional (2d) spatial input. 
In the Fourier neural layers, the normal weight matrix (usually the fully connected neural network) is replaced by the CB.
As for the projection layer, all cases adopt the CB, where 1d CNN  for the KS equation and forced KS equation, and 2d CNN for the free-decaying NS equations and forced NS equations. Except for the first Fourier neural layer, a batch norm layer is adopted before each Fourier neural layer.

Parameters for all cases are listed in the following tables:

\begin{table}[h]
    \centering
\caption{Parameters for the KS equation and forced KS equation.}
    \begin{tabular}{cc} 
    \hline
    Parameters & Value \\ 
    \hline
 Learning rate&1e-3\\ 
Learning rate scheduler& CosineAnnealingLR ($T_{max}$=70000)\\ 
Optimizer& Adam (weight decay = 1e-4)\\ 
 Activation function&GELU\\ 
    Epoch& 2000\\ 
    Batch size& 512\\ 
 Number of Fourier neural layer&4\\ 
 Mode in one Fourier neural layer&33\\ 
 Channel in one Fourier neural layer&64\\ 
\end{tabular}
\end{table}
\begin{table}[h]
    \centering
\caption{Parameters for the forced NS equations and free-decaying NS equations.}
    \begin{tabular}{cc}
    \hline
    Parameters & Value \\ 
    \hline
 Learning rate&1e-3\\ 
    Learning rate scheduler& CosineAnnealingLR ($T_{max}=70000$)\\ 
    Optimizer& Adam (weight decay = 1e-4)\\ 
 Activation function&GELU\\ 
    Epoch& 2000\\ 
    Batch size& 64\\ 
 Number of Fourier neural layer&4\\ 
 Mode in one Fourier neural layer&[48, 48]\\ 
 Channel in one Fourier neural layer&90\\ 
    
\end{tabular}

\end{table}

\newpage
\section{Section S3. Train loss and test loss}
\begin{figure}[H]
    \centering
    \includegraphics[width=0.95\textwidth]{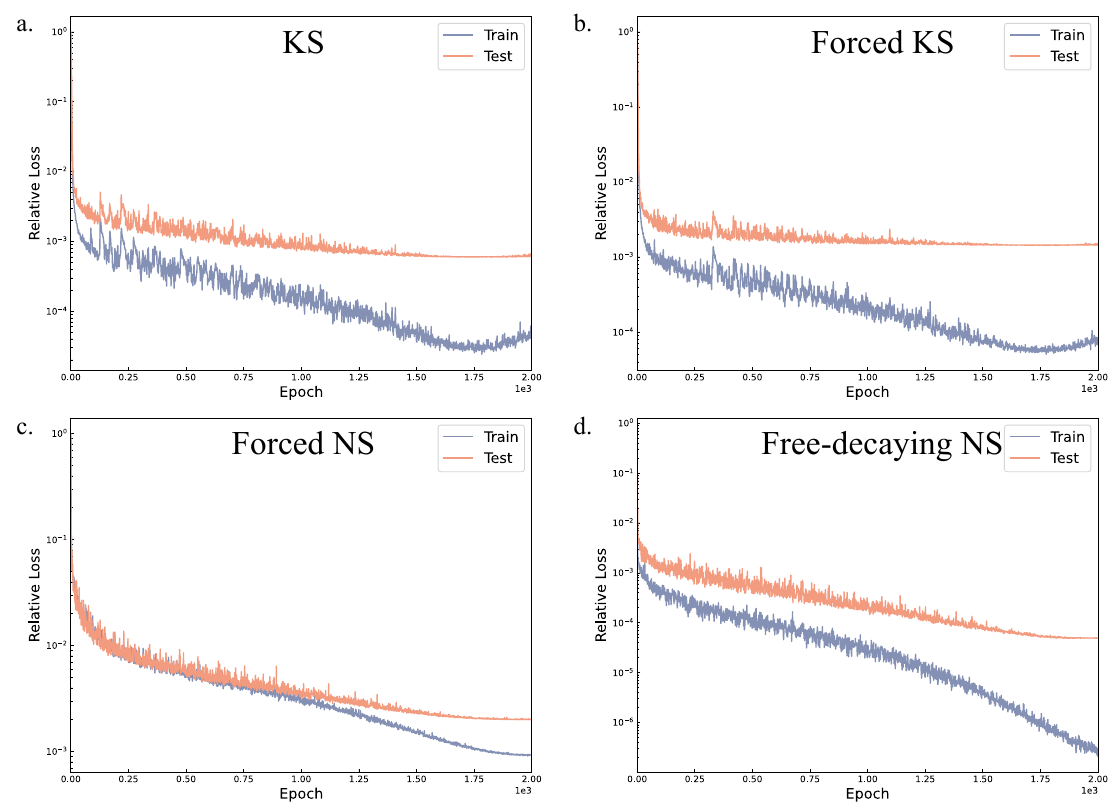}
    \caption{Illustration of the loss on the train dataset and test dataset for a. the KS equation, b. forced KS equation, c. forced NS equations, and d. free-decaying NS equations.}
    \label{loss}
\end{figure}

\newpage
\section{Section S4. Physical statistical feature}
\begin{figure}[H]
    \centering
    \includegraphics[width=0.95\textwidth]{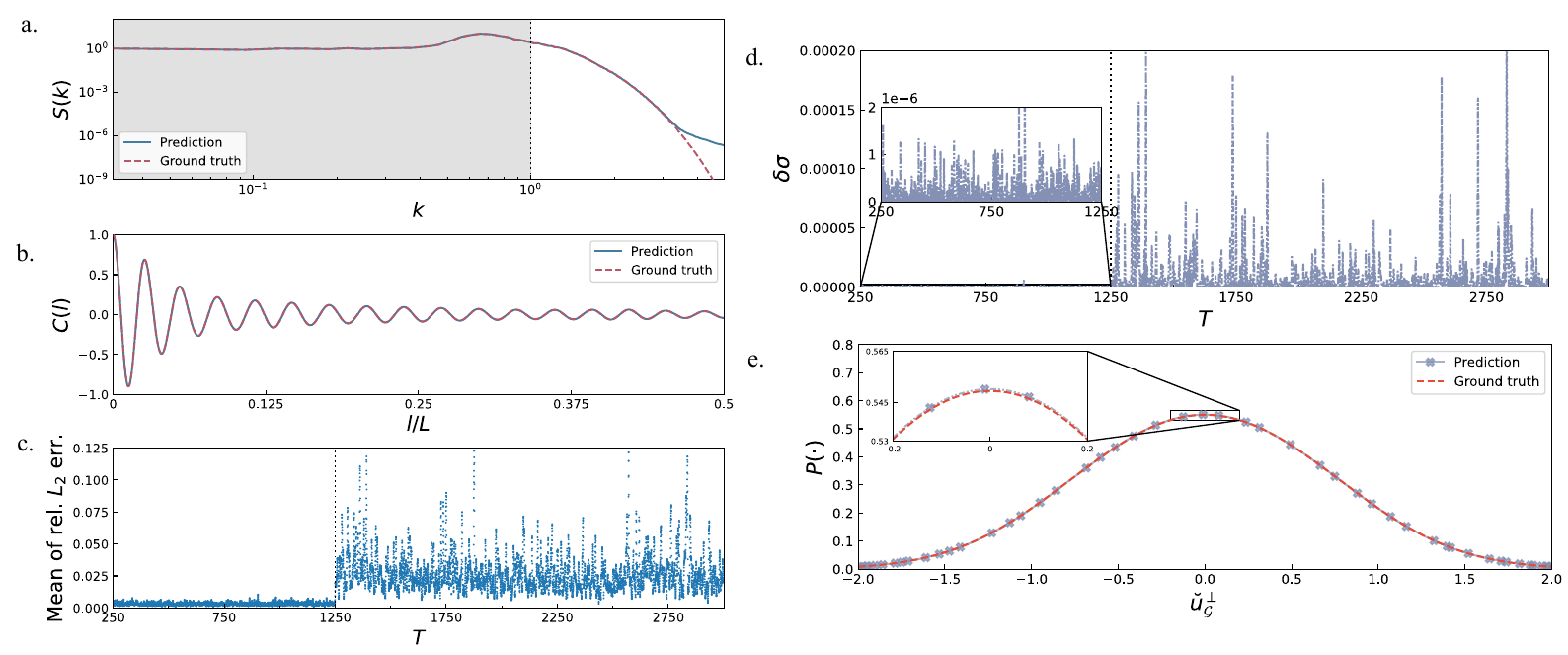}
    \caption{Illustration of comparison of physical statistical features between prediction and truth on KS equation. a. Power spectrum. b. Autocorrelation. c. Averaged relative $L_2$ error of power spectrum and autocorrelation. d. Error distance of the second moment between prediction and truth. e. PDF.}
    \label{ks_sf}
\end{figure}

\begin{figure}[H]
    \centering
    \includegraphics[width=0.95\textwidth]{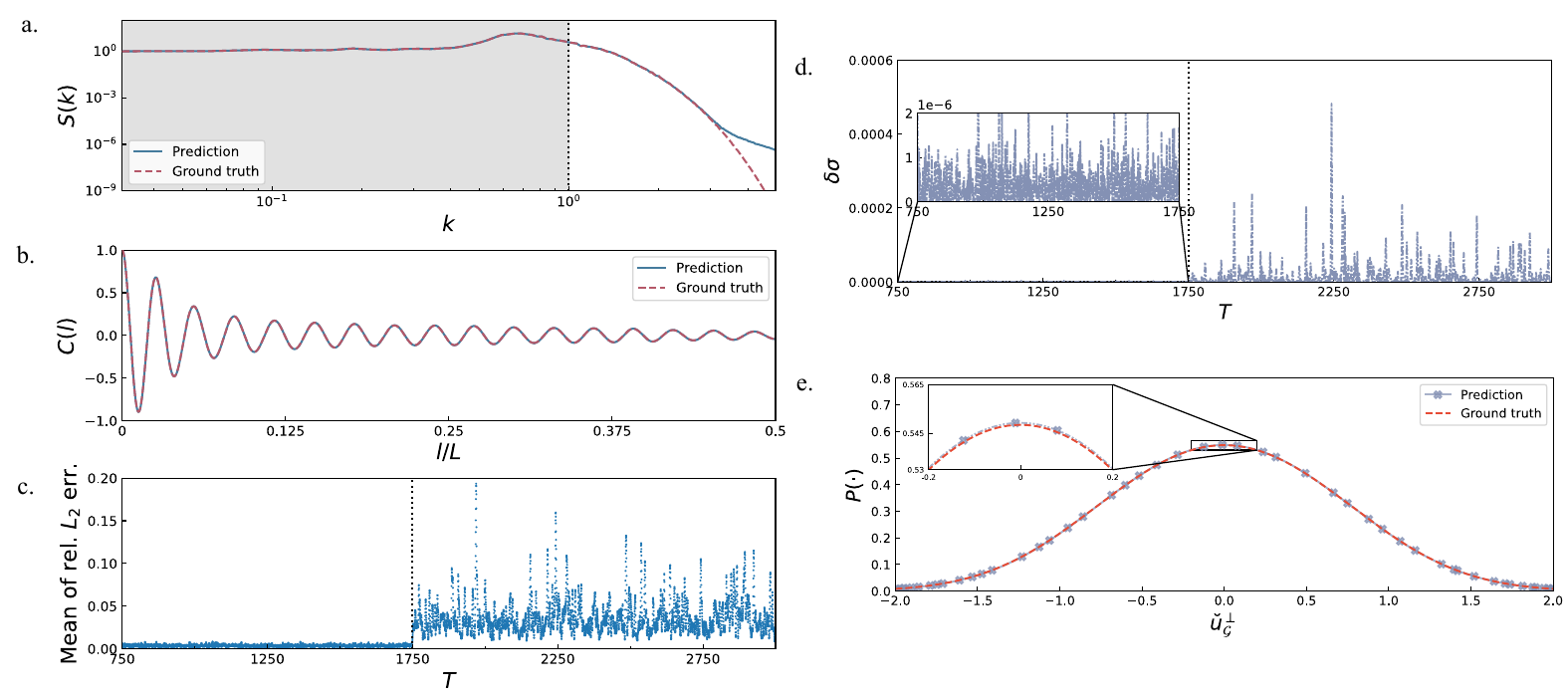}
    \caption{Illustration of comparison of physical statistical features between prediction and truth on forced KS equation. a. Power spectrum. b. Autocorrelation. c. Averaged relative $L_2$ error of power spectrum and autocorrelation. d. Error distance of the second moment between prediction and truth. e. PDF.}
    \label{fks_sf}
\end{figure}

\begin{figure}[H]
    \centering
    \includegraphics[width=0.95\textwidth]{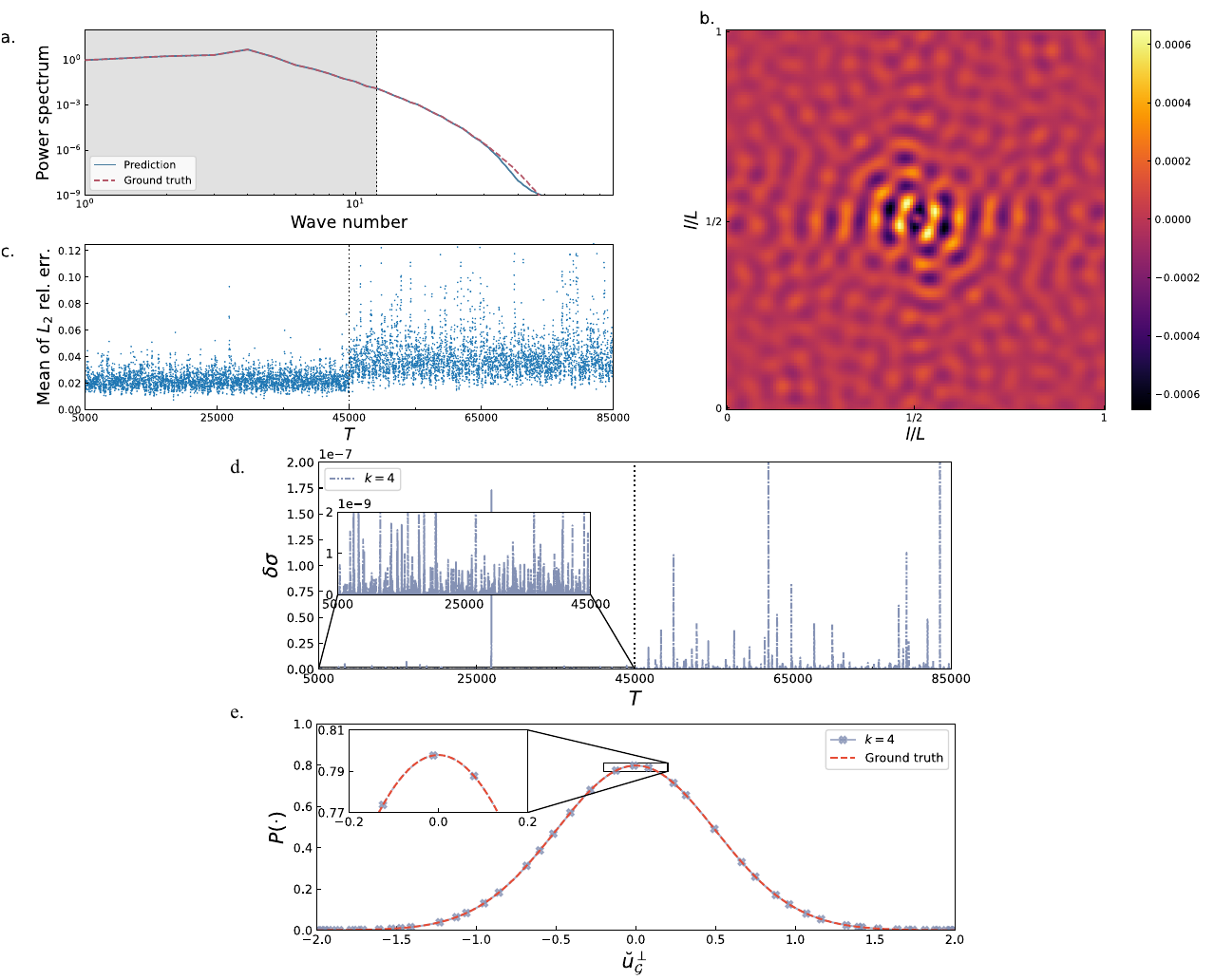}
    \caption{Illustration of comparison of physical statistical features between prediction and truth on forced NS equations. a. Power spectrum. b. Heatmap of autocorrelation. c. Averaged relative $L_2$ error of power spectrum and autocorrelation. d. Error distance of the second moment between prediction and truth. e. PDF.}
    \label{ns_sf}
\end{figure}

An analysis of statistical characteristics has been conducted for three cases, including the power spectrum, autocorrelation, second moment, and probability density function (PDF). The subsequent discussion focuses on the precise retrieval of physical statistical characteristics. A concise overview is provided, along by the depiction of the results obtained from the KS equation, forced KS equation, and forced NS equation, which are illustrated in Fig. \ref{ks_sf}, Fig. \ref{fks_sf}, and Fig. \ref{ns_sf}, respectively.

The power spectrum serves as a fundamental tool for elucidating the energy distribution across diverse frequency components within a system, offering invaluable insights into energy cascades across scales. Such analyses are pivotal in unraveling the underlying dynamics of complex systems, shedding light on both stable and unstable behaviors. 
A comparison of power spectrum $S(k)$ between predictions and ground truth is given in Fig. \ref{ks_sf}-\ref{ns_sf}a. Notably, the shadow region represents the wave number component forming the large-scale dynamics, encapsulating approximately $85.54\%$ of the energy in the KS equation, $85.58\%$ in the forced KS equation, and a substantial $99.85\%$ in the forced NS equations. 
Intriguingly, while predicted spectra closely align with the reference spectrum across the dissipation regime, slight discrepancies emerge in the high-wavenumber tail, a phenomenon commonly observed in numerical methods tackling nonlinear problems. 
 Additionally, it is noteworthy that the power spectrum $S(k)$ undergoes temporal averaging and subsequent normalization by its initial value, a practice essential for comparative analyses.

In turbulent systems, spatial coherence is assessed through the spatial autocorrelation function, offering insights into energy transfer and redistribution across scales. 
Fig. \ref{ks_sf}-\ref{fks_sf}b. shows a comparison of autocorrelation $C(l)$ between ground truth and predictions over the time horizon. It is clearly observed that all predicted $C(l)$ are in good agreement with the ground truth. Due to the 2d space of the forced NS equations, the autocorrelation error is presented in a 2d heatmap in Fig. \ref{ns_sf}b, where the maximum deviation is located at the center, with an extremely small error of up to $7e-4$. These results further indicate that ND reproduces the small-scale structures correctly by the large-scale structures. Besides, the autocorrelations $C(l)$  are averaged on time and then are also normalized by its own very first value.

Especially, the power spectrum $S(j)$ and the autocorrelation function $C(l)$ form a Fourier transform pair, which is also known as the Wiener–Khinchin theorem. It means that the power spectrum difference between prediction and ground truth in the high-wavenumber region is aligned with the autocorrelation difference on the half of spatial lag number, and the difference over the time horizon is shown in Fig. \ref{ks_sf}-\ref{ns_sf}c, where very accurate results are obtained. It achieves $0.47\%$ in the relative $L_2$ errors on training dataset and $2.44\%$ in the relative $L_2$ errors for KS euqation, $0.69\%$ and $3.79\%$ for forced KS euqation, $2.83\%$ and $4.47\%$ for forced KS euqations.

To capture the essence of complexity and nonlinearity, the second moment is usually adopted to quantify the extent of variability within a system, allowing one to learn the uncertainty in nonlinear complex systems. 
In Fig. \ref{ks_sf}-\ref{ns_sf}d, the error distance $\delta \sigma$ of the second moment between the truth and prediction is shown. It is clear that much lower error is achieved, which
demonstrates that high-fidelity reconstruction is obtained by ND. 
The second moment $ \sigma_{\mathcal{G}}$ of the  small-scale slaved dynamics $\breve{u}_{\mathcal{G}, i}^{\perp}$ is defined by 
\begin{equation}
    \sigma_{\mathcal{G}} = \left(\overline{(\breve{u}_{\mathcal{G}, i}^{\perp})^2}-\overline{\breve{u}_{\mathcal{G}, i}^{\perp}}^2\right)^{1/2},
\end{equation}
where the bar atop variable indicates that the variable is averaged on the time and space. The second moment $\sigma$ of ground truth can be defined in the same way. Then the error distance $\delta \sigma$ is defined by
\begin{equation}
    \delta \sigma = \sigma_{\mathcal{G}} - \sigma.
\end{equation}

For small-scale dynamics, where the intricacies of interactions and fluctuations are pronounced, the PDF allows for the prediction and analysis of rare events and extreme conditions that may not be evident through mean-field approaches. Understanding the tails of the distribution—representing low-probability, high-impact events—is as crucial as the average outcomes. 
In order to get PDF of small-scale slaved dynamics, we estimate it by using the Kernel Density Estimation method. The PDF can be calculated by the following formula
\begin{equation}
    P = \frac{1}{nh} \sum_{i=1}^n \mathcal{K}\left (\frac{\breve{u}_{\mathcal{G}}^{\perp}-\breve{u}_{\mathcal{G}, i}^{\perp}}{h}\right)
\end{equation}
where $n$ indicates the number of samples including time and space, $\mathcal{K} (\cdot)$ indicates the kernel function and $h$ indicates the bandwidth which is the parameter for smoothing the PDF distribution. Gaussian kernel with $h=0.5$ is adopted to estimate the PDF.  To achieve a fast estimation of PDF, the number of samples is reduced to $1/10$ on the time horizon and the spatial direction for KS equation and forced KS equation, $1/7$ for forced NS equations. 
As depicted in Fig. \ref{ks_sf}e, the PDF estimated by ND's prediction coincides with the one by truth, showcasing ND's accurate reconstruction of the small-scale vorticity distribution.

\newpage
\section{Section S5. Insight into complexity reduction by the investigation of scale decomposition}
Here the investigations of scale decomposition for three cases are performed to shed light on the complexity reduction in complex systems. It provides a new perspective for us to seek the essential components in complex systems and even possesses the potential to instruct the construction and validation of the reduced order model in the future.

\begin{figure}[ht]
    \centering
    \includegraphics[width=0.95\textwidth]{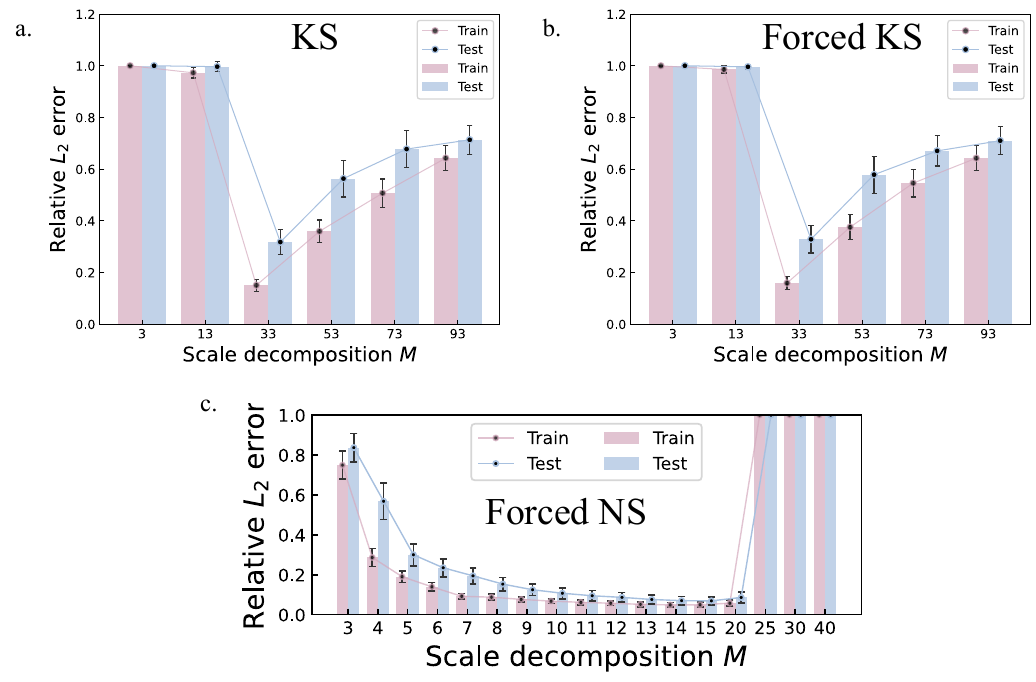}
    \caption{Illustration of investigations of temporal-mixing step for a. the KS equation, b. forced KS equation and c. forced NS equations. The height of bar represents the mean and the error bar represents the variance in dataset, where pink indicates training dataset and blue test dataset.}
    \label{scale_decomposition}
\end{figure}

To facilitate the comparison between different scale decomposition, the simplicity of ND, with 16 modes and 32 channels in one Fourier layer for the KS equation and forced KS equation and 24 modes and 45 channels for the forced NS equations, is trained in 200 epochs. 
As observed in Fig. \ref{scale_decomposition},  the relative $L_2$ error decreases as scale decomposition parameter increases until reaching a critical threshold, approximately \$33\$ for both the KS equation and forced KS equation, and approximately \$20\$ for the forced NS equations. Beyond this critical point, a reversal of this trend occurs, characterized by an increase in the relative $L_2$ error with further increments in the scale decomposition parameter. 
The identification of the scale decomposition parameter associated with optimal performance offers valuable insights into achieving complexity reduction in intricate systems. Further elaboration on this aspect is provided in Section S6.

\newpage
\section{Section S6. The influence of the large-scale dynamics}
To extend more general scenarios with \textbf{an arbitrary number of modes for large-scale dynamics}, a compensation operator is necessary for compensating the missing or exceeding dynamics. 
According to the investigation of the number of scale decomposition modes $M$ in Section S7, there always exists the best scale decomposition for constructing the approximate inertia manifold to transform the large-scale dynamics to the small-scale ones. As expected, the construction of the inertia manifold $\mathscr{M}$ depends on the assumption that small-scale dynamics $q_m$ is small compared with the large-scale dynamics $u$. 

When the large-scale dynamics consist of fewer modes and the small-scale dynamics consist of more modes in turn, small-scale dynamics $q_m$ grow and the nonlinear parts involved $q_m$, i.e. $\mathscr{N}(q_m, q_m)$, $\mathscr{N}(q_m, p_m)$ and $\mathscr{N}(p_m, q_m)$, also grow. In comparison with $\mathscr{N}(p_m, p_m)$, then they can not be neglected, leading to the mixture of large-scale and small-scale dynamics and the absence of explicit graph mapping between them.

When the large-scale dynamics consist of more modes and the small-scale dynamics consist of fewer modes in turn, small-scale dynamics $q_m$ decrease and all nonlinear parts involved $q_m$ can be neglected. Eq. (3) in the main text becomes the form of
\begin{equation}
    \mathscr{L}q_m(t)+\mathscr{Q}_Mf=0.
\end{equation}
This ultimately leads to the absence of explicit graph mapping between them. They account for the degenerated precision in the number of scale decomposition modes that are far away from the critical number. Therefore, based on the existence of explicit graph mapping, it is reasonable to believe that the scale decomposition with best performance is exactly the optimal separating point. Large-scale dynamics, induced by this optimal point, include all necessary information to accurately predict system's long-term behaviour.

As far, vanilla ND can not handle this implicit modelling problem with satisfactory performance. Based on the optimal graph mapping formed by the optimal separating point above, we here present the construction of a compensation operator as a hopeful solution. Similar to the non-uniform meshes problem, a compensation operator can be incorporated atop of ND but its inversion does not need to be established independently since the equivalent of the exceeding (missing) part in large-scale and the missing (exceeding) part in small-scale.

\newpage
\section{Section S7. Necessity of the temporal-mixing step for predicting small-scale dynamics}
Here, an examination of different numbers of temporal-mixing steps is conducted across three cases. This investigation serves to underscore the significance of incorporating temporal dependence in both modeling and computation, particularly concerning the dynamics of small scales.

\begin{figure}[H]
    \centering
    \includegraphics[width=0.95\textwidth]{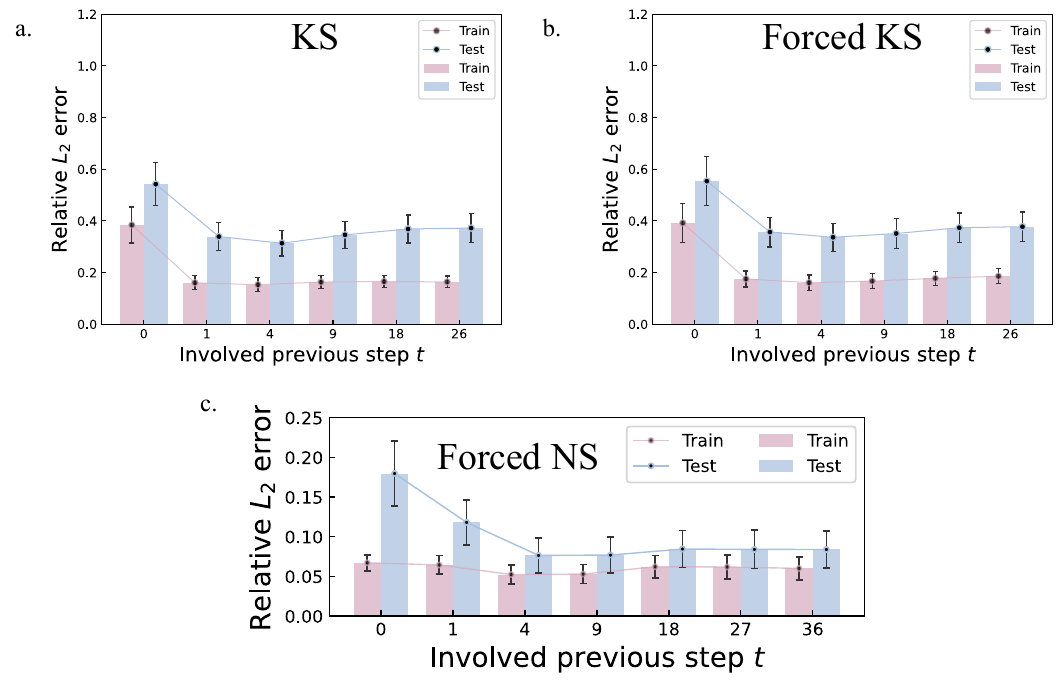}
    \caption{Illustration of investigations of temporal-mixing step for a. the KS equation, b. forced KS equation and c. forced NS equations. The height of bar represents the mean and the error bar represents the variance in dataset, where pink indicates training dataset and blue test dataset.}
    \label{temporal_mixing}
\end{figure}

To facilitate the comparison between different numbers of temporal-mixing step, the simplicity of ND, with 16 modes and 32 channels in one Fourier layer for the KS equation and forced KS equation and 24 modes and 45 channels for the forced NS equations, is trained in 200 epochs. 
As observed in Fig. \ref{temporal_mixing}, relative $L_2$ errors on multiple temporal steps are less than the ones on a single temporal step on both training dataset and test dataset, and only a tiny difference between multiple temporal steps exists. Notably, the temporal mixing-induced error decrease on training dataset of the forced NS equation is not as obvious as the other two cases. It may be caused by a large time resolution that enables unconnected between two adjacent steps, in other words, dynamics in two adjacent steps are completely independent such that ND can not capture the context information between them. Although the error decrease seems slight, ND still maintains a promising error reduction on test dataset, highlighting the excellent effect of the temporal-mixing step on generalization. Also, this observation reminds us that the time resolution plays an important role in the temporal-mixing step for enhancing predicting small-scale dynamics. It will be explored deeply in future work.

\bibliography{sn-bibliography}
